\documentclass[letterpaper]{JHEP3}
\usepackage{amssymb,amsfonts}
\usepackage{cite}
\usepackage{epsfig}

\newcommand{\CA}{{\cal A}}
\newcommand{\CD}{{\cal D}}
\newcommand{\CF}{{\cal F}}
\newcommand{\CG}{{\cal G}}

\newcommand{\CL}{{\cal L}}
\newcommand{\CM}{{\cal M}}
\newcommand{\CN}{{\cal N}}
\newcommand{\CO}{{\cal O}}
\newcommand{\CR}{{\cal R}}

\newcommand{\CV}{{\cal V}}
\newcommand{\CW}{{\cal W}}

\newcommand{\CZ}{{\cal Z}}

\newcommand{\bx}{{\bf x}}

\newcommand{\FA}{{\mathfrak A}}
\newcommand{\FQ}{{\cal Q}}
\newcommand{\FS}{{\cal S}}
\newcommand{\p}{\partial}
\renewcommand{\bar}[1]{\overline{#1}}
\renewcommand{\tilde}[1]{\widetilde{#1}}
\newcommand{\be}{\begin{equation}}
\newcommand{\ee}{\end{equation}}
\newcommand{\bea}{\begin{eqnarray}}
\newcommand{\eea}{\end{eqnarray}}

\newcommand{\ie}{{\it i.e.}}
\newcommand{\eg}{{\it e.g.}}

\usepackage{graphicx}
\usepackage{latexsym}
\newcommand\x\bx

\newcommand\D\Delta
\newcommand\n\nabla
\mathchardef\mhyphen="2D
\newcommand\deltaD{\delta^{\ }_\CD}
\newcommand\fin{{\rm fin}}
\newcommand\Sonshell{S_{\rm on\mhyphen shell}}
\title{Conformal Lifshitz Gravity from Holography}
\author{Tom Griffin, Petr Ho\v{r}ava and Charles M. Melby-$\!$Thompson
\\
Berkeley Center for Theoretical Physics and Department of Physics\\
University of California, Berkeley, CA, 94720-7300\\
and\\
Theoretical Physics Group, Lawrence Berkeley National Laboratory\\
Berkeley, CA 94720-8162, USA}
\abstract{We show that holographic renormalization of relativistic gravity in 
asymptotically Lifshitz spacetimes naturally reproduces the structure of 
gravity with anisotropic scaling: The holographic counterterms induced near 
anisotropic infinity take the form of the action for gravity at a Lifshitz 
point, with the appropriate value of the dynamical critical exponent $z$.  
In the particular case of $3+1$ bulk dimensions and $z=2$ asymptotic scaling 
near infinity, we find a logarithmic counterterm, related to anisotropic Weyl 
anomaly of the dual CFT, and show that this counterterm reproduces precisely 
the action of conformal gravity at a $z=2$ Lifshitz point in $2+1$ dimensions, 
which enjoys anisotropic local Weyl invariance and satisfies the detailed 
balance condition.  We explain how the detailed balance is a consequence of 
relations among holographic counterterms, and point out that a similar 
relation holds in the relativistic case of holography in $AdS_5$.  
Upon analytic continuation, analogous to the relativistic case studied 
recently by Maldacena, the action of conformal gravity at the 
$z=2$ Lifshitz point features in the ground-state wavefunction of a 
gravitational system with an interesting type of spatial anisotropy.}
\begin{document}
% \maketitle
%%%%%%%%%%%%%%%%%%%%%%%%%%%%%%%%%%%%%%%%%%%%%%%%%%%%%%%%%%%%%%%%%%%%%%
\section{Introduction}

The possibility that gravity may exhibit multicritical 
behavior with Lifshitz-like anisotropic scaling at short distances 
\cite{mqc,lif} has attracted a lot of attention recently (see, \eg , 
\cite{grx,revmv,revsm} for some reviews).  Such multicritical quantum gravity 
can be formulated as a field theory of the fluctuating spacetime metric, 
characterized by scaling which is anisotropic between time and space,
\be
\label{ascaling}
t\mapsto b^zt,\qquad\bx\mapsto b\bx,\qquad
\ee
with dynamical exponent $z$.    

When $z$ equals the number of spatial dimensions $D$, several interesting 
things happen:  First, the theory becomes power-counting renormalizable, 
when we allow all terms compatible with the gauge symmetries in the action.  
In addition, the effective spectral dimension of spacetime flows 
to two at short distances, in accord with the lattice results first obtained 
in the causal dynamical triangluations approach to quantum gravity in 
\cite{ajld,ajl,dario}, and independently confirmed recently in \cite{wdavis}.  
Moreover, when $z=D$, one can further restrict the classical action by 
requiring its invariance under the local version of the rigid anisotropic 
scaling, which acts on the spacetime metric via anisotropic Weyl 
transformations.  This leads to an anisotropic version of conformal 
gravity \cite{mqc}.  

While such multicritical gravity models may not need string theory for a UV 
completion, it is still natural to ask whether they can be engineered from 
string theory.  Indeed, it seems likely that any mathematically consistent 
theory of gravity should have a role to play in the bigger scheme of strings.  
Here we present one specific construction in which the action of multicritical 
gravity with anisotropic scaling appears naturally from string theory and 
AdS/CFT correspondence: The holographic renormalization of spacetimes which 
are asymptotically Lifshitz, or in other words, dual to nonrelativistic field 
theories with Lifshitz scaling.  

In recent years, the AdS/CFT correspondence has been extended to 
spacetimes which are asymptotically non-relativistic, 
with the hope of providing new techniques for understanding strongly coupled 
condensed matter systems using dualities originating in string theory (see, 
\eg , \cite{revsh,revmcg,revss,vhmr} for recent reviews of this program).  
Such asymptotically non-relativistic spacetimes fall into two classes: 
Either they approach Schr\"odinger symmetries \cite{son,bmcg}, or they exhibit 
Lifshitz-type scaling \cite{klm}.  In both cases, Penrose's standard 
definition of conformal infinity (see, \eg, \cite{penrin}) gives results 
which are puzzling and appear inconsistent with the expectations based on 
gauge-gravity duality.  
It turns out that for spacetimes which carry an asymptotic foliation 
structure, a natural generalization of Penrose's notion of conformal infinity 
to asymptotically anisotropic spacetimes exists \cite{aci}, and it reproduces 
features expected from holography.         

This clearer picture \cite{aci} of the asymptotic structure of Lifshitz 
spacetimes allows us to perform holographic renormalization, 
study the precise structure of holographic counterterms, and compare the 
results to the relativistic case.  This is the goal of the present paper.  
We focus mainly on the case of $3+1$ bulk dimensions, in particular with 
the $z=2$ scaling.  In this case, we find a logarithmic counterterm, which 
takes the form of the action of the $z=2$ conformal multicritical gravity in 
$2+1$ dimensions.  In relativistic AdS/CFT, logarithmic gravitational 
counterterms appear only when the dimension $d$ of the boundary 
is even.  In that circumstance, they take the form of the Weyl anomaly 
\cite{mans} (see \cite{weylduff} for a review of the Weyl anomaly).  For 
example, in the maximally supersymmetric case in $d=4$, the anomaly is given 
by the action of conformal supergravity \cite{bgmr} (see, \eg , \cite{confsg} 
for a review of the early history of conformal supergravity).  
In Lifshitz spacetimes, the logarithmic counterterms -- if and when they 
appear -- should be related to the little-studied nonrelativistic Weyl 
anomaly (see \cite{theisen} for some early results on the Weyl anomaly at 
$z=3$ in $d=4$, and \cite{ibdl,ib} for a detailed discussion of axial 
anomalies in Lifshitz theories).  In Appendix~\ref{appwano}, we briefly 
discuss the cohomological structure of the $z=2$ anisotropic Weyl anomaly in 
$2+1$ dimensions, and (modulo total derivatives) find two independent terms 
that can appear in the 
action.  However, perhaps surprisingly, the action that we obtain in the 
logarithmic counterterm turns out to satisfy the additional condition of 
detailed balance, which reduces the number of independent terms to one.  
We show how this condition is implied by the machinery of holographic 
renormalization, which relates the logarithmic counterterm to another, 
quadratic counterterm.  

The techniques of holographic renormalization in asymptotically $AdS$ 
spacetimes can also be usefully applied, upon appropriate Wick rotation, 
to asymptotically de~Sitter geometries \cite{revsk}, leading to results about 
the ground-state wavefunction of the universe at superhorizon scales 
\cite{malda} (see also \cite{maldai,harlow}, and the series of papers 
\cite{mcf1,mcf2,mcf3}).  
Since holography in asymptotically Lifshitz spacetimes parallels closely 
the case of $AdS$, it is natural to perform the corresponding Wick rotation, 
obtain a candidate ground-state wavefunction, and ask what kind of 
gravitational system this wavefunction describes.  In the case of $z=2$ and 
bulk $3+1$ dimensions, we show that this wavefunction corresponds to a 
spatially anisotropic gravity theory with an interesting form of 
ultralocality.  

Our discussion in the bulk of the paper mostly focuses on the case of 
$3+1$ bulk dimensions, in particular with $z=2$.  However, after summarizing 
our conventions and notation in Appendix~\ref{appnot}, we present our 
detailed calculations also for general $D$ and $z$ in an extensive 
Appendix~\ref{apphre}, with the hope that the inquisitive reader may find 
the results useful.  

\section{Gravity at a Lifshitz Point}
\label{lifgrav}

In this section, we briefly review some features of gravity with anisotropic 
scaling, concentrating on aspects relevant for the main points of this paper.  

The theory can be formulated in the general number of $d=D+1$ spacetime 
dimensions.  Since the spacetime manifold $M$ is assumed to carry a preferred 
foliation structure $\CF$, consisting of codimension-one leaves $\Sigma$ of 
constant absolute time, it is natural to use nonrelativistic coordinates $t$ 
and $\bx\equiv\{x^i, i=1,\ldots D\}$, adapted to the foliation.  In such 
coordinates, the theory is then described by specifying its fields and its 
symmetries.  The gravity field multiplet consists of fields 
\be
\label{mult}
N,\quad N_i,\quad \gamma_{ij},   
\ee
familiar from the ADM decomposition of the relativistic 
metric on spacetime:  $N$ is the lapse function, $N_i$ the shift vector, and 
$\gamma_{ij}$ the spatial metric on the leaves $\Sigma$.  Ocassionally, we 
will refer to the set of fields (\ref{mult}) as the ``metric multiplet,'' to 
indicate that the ADM split is not just a convenience, but a reflection of the 
preferred foliation structure of spacetime.  To think of the spacetime metric 
as one irreducible field would be misleading in the context of Lifshitz 
gravity.  

In the simplest version of the theory, the gauge symmetries are given  
by those spacetime diffeomorphisms that preserve the preferred foliation 
$\CF$.  Such foliation-preserving diffeomorphisms ${\rm Diff}(M,\CF)$, 
generated by 
\be
\label{fdiff}
\delta t=f(t),\qquad \delta x^i=\xi^i(\bx,t),
\ee
contain one fewer gauge symmetry per spacetime point than the symmetries of 
general relativity.  Theories of gravity with anisotropic scaling whose 
symmetries are as large as those of general relativity can be constructed 
\cite{gen}, but they will stay outside of the scope of the present paper.  

The action respecting the symmetries of ${\rm Diff}(M,\CF)$ consists of a 
kinetic term,
\be
S_K=\frac{1}{\kappa^2}\int dt\,d^D\bx\,\sqrt{\gamma}\,N\left(\hat{K}_{ij}
\hat{K}^{ij}-\lambda \hat{K}^2\right),
\ee
where 
\be
\hat{K}_{ij}=\frac{1}{2N}\left(\partial_t\gamma_{ij}-\nabla_iN_j-\nabla_jN_i
\right)
\ee
is the extrinsic curvature of $\Sigma$, $\hat K\equiv\hat K_i^i$, and 
$\lambda$ is dimensionless coupling constant; and a potential term
\be
S_\CV=\frac{1}{\kappa^2}\int dt\,d^D\bx\,\sqrt{\gamma}\,N\,\CV,
\ee
with $\CV$ a scalar function built out of the spatial Riemann tensor and its 
covariant spatial derivatives, but independent of the time derivatives of all 
fields.  Among these terms, the spatial scalar curvature $R$ and the constant 
term dominate at long distances, while terms of scaling dimension $2z$ 
take over the dynamics at shortest scales.  

In higher dimensions, and for higher values of $z$, the number of available 
relevant and marginal terms that can appear in $\CV$ proliferates quickly.  
One can further limit the independent terms by imposing an additional 
symmetry.  For example, one can impose the {\it detailed balance 
condition\/} \cite{mqc,lif}.  
This condition means that $\CV$ is constructed from an auxiliary local action 
$\CW$ in $D$ Euclidean dimensions, as the sum of squares of the $\CW$ 
equations of motion:
\be
\CV=\CG_{ijk\ell}\frac{\delta\CW}{\delta\gamma_{ij}}\frac{\delta\CW}{\delta
\gamma_{k\ell}},
\ee 
with an appropriately chosen non-derivative DeWitt metric tensor 
$\CG_{ijk\ell}$.  This condition -- inspired by the theory of 
non-equilibrium systems -- has a straightforward generalization in the 
presence of matter.   When the theory is in detailed balance, the number of 
independent couplings in $\CV$ reduces to the number of independent couplings 
in $\CW$.  

Note also that since the lapse function is the gauge field associated with 
the time reparametrization symmetry, it can be naturally restricted to be 
a function of time only; this leads to the projectable version of the theory.  
In that case, the potential term $\CV$ is a local function of the Riemann 
tensor of the spatial metric $\gamma_{ij}$, and its covariant derivatives.  
It is also possible to relax the projectability condition, and allow $N$ to 
be a spacetime-dependent field; this yields the non-projectable version 
\cite{mqc,lif,bpstwo,bpsthree,grx}.  
In the non-projectable theory, there is one additional ingredient that 
can be used to build the potential term $\CV$: the spatial vector field 
$\nabla_iN/N$.  

\subsection{Anisotropic Weyl Transformations}
\label{awt}

Under certain circumstances, we can impose additional gauge symmetries to 
further constrain the classical action of gravity with anisotropic scaling.  
When $z=D$, one can thus require invariance under a local version of the 
anisotropic scaling (\ref{ascaling}), which acts on the metric multiplet by 
anisotropic Weyl transformations 
\be
\label{aweyl}
N \mapsto e^{z\omega}N
\qquad
N_i \mapsto e^{2\omega}N_i
\qquad
\gamma_{ij} \mapsto e^{2\omega}\gamma_{ij},
\ee
with an arbitrary local function $\omega(t,\bx)$.  We denote the group of 
anisotropic Weyl transformations (\ref{aweyl}) with dynamical exponent $z$ by 
${\rm Weyl}_z(M,\CF)$.  Crucially, this group extends the group of foliation 
preserving diffeomorphisms into a semi-direct product \cite{mqc,aci} 
\be
\label{semiwd}
{\rm Weyl}_z(M,\CF)\rtimes{\rm Diff}(M,\CF).
\ee
Indeed, the commutator between an infinitesimal foliation-preserving 
diffeomorphism $\delta_{(f,\ \xi^i)}$ of (\ref{fdiff}) and an infinitesimal 
generator $\delta_\omega$ of the anisotropic Weyl transformation 
(\ref{aweyl}) yields another infinitesimal anisotropic Weyl transformation, 
\be
[\delta_\omega,\delta_{(f,\ \xi^i)}]=\delta_{f\p_t\omega+\xi^i\p_i\omega},
\ee
with the same fixed -- but otherwise arbitrary -- value of $z$.  On the other 
hand, had we tried to extend ${\rm Diff}(M,\CF)$ 
into the full spacetime diffeomorphism group, the closure of the symmetries 
would have forced the relativistic scaling with $z=1$.  Thus, anisotropic 
Weyl symmetry is only possible when we relax the spacetime diffeomorphism 
symmetry to the symmetries of the preferred foliation $\CF$.  

Since ${\rm Weyl}_z(M,\CF)$ acts on $N$ by a spacetime-dependent gauge 
transformation (\ref{aweyl}), $N$ itself must be a spacetime-dependent field, 
hence it cannot satisfy the projectability condition.   This suggests that 
the natural environment for conformal gravity with anisotropic Weyl invariance 
is the non-projectable version of the theory.%
\footnote{One might consider restricting the Weyl invariant combination 
$\tilde N\equiv N/\sqrt{\gamma}$ to be a function of time only \cite{mqc}; 
we leave the study of such a ``conformally projectable'' theory outside of the 
scope of the present paper.}

\subsection{$z=2$ Conformal Gravity in $2+1$ Dimensions}

Insisting on the additional symmetries (\ref{aweyl}) implies that the coupling 
constant $\lambda$ must take a fixed value, $\lambda=1/D$.  We will refer to 
it as the ``conformal value'' of $\lambda$.  In this paper, we will be mainly 
interested in the case of $D=2$, which requires $z=2$ and the unique 
kinetic term
\be
\label{confk}
S_K=\frac{1}{\kappa^2}\int dt\,d^2\bx\,\sqrt{\gamma}\,N\left(\hat{K}_{ij}
\hat{K}^{ij}-\frac{1}{2}\hat{K}^2\right).  
\ee
One can easily check that this term is indeed invariant under (\ref{aweyl}).  

The potential term $\CV$ is also strongly constrained by the 
condition of anisotropic Weyl invariance (\ref{aweyl}).  In $D=2$, where 
the Riemann tensor of the spatial metric reduces to the Ricci scalar $\hat R$, 
there is only one term that can appear in $\CV$:%
\footnote{Throughout the paper, we use the compact notation 
$\nabla^2N\equiv\nabla i\nabla^iN$ and $(\nabla N)^2\equiv\nabla_i 
N\nabla^iN$.}
\be
\label{uniquev}
S_\CV=\frac{1}{\kappa_\CV^2}\int dt\,d^2\bx\,\sqrt{\gamma}\,N\left\{\hat{R}
+\frac{\nabla^2 N}{N}-\left(\frac{\nabla N}{N}\right)^2 \right\}^2.
\ee
This term is also invariant under (\ref{aweyl}), but it does not satisfy 
the detailed balance condition:  There is no {\it local\/} action that would 
yield this term as the sum of squares of its equations of motion.%  
\footnote{However, as was discussed in \cite{mqc}, one {\it can\/} 
get $\CV\sim R^2$ by squaring the equation of motion of a nonlocal action: 
the Polyakov conformal anomaly action 
$\int d^2\bx\sqrt{\gamma}R\frac{1}{\nabla^2}R$.}
Thus, pure $z=2$ conformal gravity in $2+1$ dimensions with detailed balance 
has no potential term.  

This conformal $z=2$ gravity in $2+1$ dimensions can be coupled to scalars 
$X^a(t,\bx)$.  Anisotropic Weyl invariance of the classical action will be 
preserved when we assign scaling dimension zero to $X^a$.  The kinetic term 
becomes
\be
S_K=\frac{1}{\kappa^2}\int dt\,d^2\bx\,\sqrt{\gamma}\,N\left(\hat{K}_{ij}
\hat{K}^{ij}-\frac{1}{2}\hat{K}^2\right) + \frac{1}{2}\int dt\,d^2\bx\,
\frac{\sqrt{\gamma}}{N}\left(\partial_t X^a-N^i\p_iX^a\right)^2.
\ee
Even under the condition of detailed balance, this coupled theory develops 
a nontrivial potential.  There is a unique potential term compatible both 
with anisotropic conformal invariance and the detailed balance condition, 
\be
\label{dbpt}
S_\CV=\int dt\,d^2\bx\,\sqrt{\gamma}\,N\left\{(\nabla^2 X^a)^2
+\frac{\kappa^2}{2}\left(\p_iX^a\p_jX^a
-\frac{1}{2}\gamma_{ij}\p^kX^a\p_kX^a\right)^2\right\}.
\ee
This theory, of $z=2$ conformal gravity coupled to scalars and satisfying 
the detailed balance condition, first appeared in \cite{mqc} as the worldvolume 
action of ``membranes at quantum criticality,'' whose ground-state 
wavefunction on Riemann surface $\Sigma$ reproduces the partition function of 
the critical bosonic string on $\Sigma$.  The Euclidean action in $D=2$ 
dimensions which yields (\ref{dbpt}) via the detailed balance construction 
is simply given by the action familiar from the critical string,
\be
\label{stract}
\CW=\frac{1}{2}\int d^2\bx\,\sqrt{\gamma}\,\gamma^{ij}\p_iX^a\p_jX^a.
\ee
We recognize the first term in (\ref{dbpt}) as the square of the $X^a$ 
equation of motion, and the second term as the square of the energy-momentum 
tensor obtained from the $\gamma_{ij}$ variation of (\ref{stract}).  

\section{Holography in Asymptotically Lifshitz Spacetimes}

The metric of the Lifshitz spacetime $\CM$ in $D+2$ dimensions, 
\be
\label{lifmet}
ds^2=-r^{2z}dt^2+r^2d\bx^2+\frac{dr^2}{r^2}, 
\ee
is designed so that its isometries match the expected conformal 
symmetries of Lifshitz field theory with dynamical exponent $z$.  This 
geometry, and its various cousins, appears as a solution in several effective 
theories, such as the theory considered in \cite{taylor} in which bulk 
Einstein gravity is coupled to a massive vector, and more recently also in a 
variety of constructions obtained from string theory \cite{st1,st2,st3,st4}.  

In this section, 
we first discuss some general features of holography and asymptotic structure 
of Lifshitz spacetime, which are universal and independent of the precise 
model.  Then,  in Section~\ref{sechr}, we work -- for specificity -- in the 
effective bulk theory of relativistic gravity coupled to a massive vector, 
first without additional matter, and then coupled to bulk scalars.  Even 
though our detailed results will depend of the chosen effective theory, we 
believe that our conclusions are largely universal and generalizable 
straightforwardly to other embeddings of Lifshitz spacetimes.

\subsection{Anisotropic Conformal Infinity}

The notion of conformal infinity plays a central role in general relativity 
\cite{hawkinge,penrin}.  It is constructed by mapping the original metric%
\footnote{We are using Penrose's ``abstract index'' notation.}
$G_{\mu\nu}$ on a manifold $\CM$ via a smooth conformal Weyl transformation to 
\be
\label{eqnom}
\tilde G_{\mu\nu}=\Omega^2(x)G_{\mu\nu},
\ee
such that the rescaled metric $\tilde G_{\mu\nu}$ is extendible to a larger 
manifold $\tilde\CM$, which contains the closure $\bar\CM$ of $\CM$ as a 
closed submanifold.  The idea is to define the conformal infinity of $\CM$ to 
be the set $\bar\CM\setminus\CM$.  The scaling factor $\Omega$ must extend to 
$\tilde\CM$ and satisfy certain regularity conditions at $\bar\CM\setminus\CM$ 
(the most essential being that it should have a single zero there and that its 
exterior derivative should be nonzero), but is otherwise arbitrary.  A change 
from one permissible scaling factor to another is interpreted as a conformal 
transformation at $\bar\CM\setminus\CM$:  Hence, conformal infinity carries a 
naturally defined preferred conformal structure.  

This notion of conformal infinity allows one to define precisely, and 
in a coordinate-independent 
way, the notion of an event horizon (and hence the notion of black holes), as 
the boundary of the causal past of the future infinity.  Moreover, it allows 
us to define precisely the concept of spacetimes which ``asymptotically 
approach'' a chosen vacuum solution ``at infinity.''  In the example of 
anti-de~Sitter spacetime, this picture is naturally compatible with the 
physical ideas of holography:  The conformal infinity of $AdS$ is of 
codimension one, and carries the natural conformal structure induced from the 
asymptotic isometries of the bulk, just as predicted by the holographic 
dictionary.  

In contrast, the intuition of holographic renormalization in Lifshitz and 
Schr\"odinger spacetimes clashes with this classic notion of conformal 
infinity as defined by Penrose: For the Lifshitz spacetime (\ref{lifmet}) with 
$z>1$, the relativistic conformal infinity is of dimension one for any $D$, 
and it does not inherit the conformal structure expected from the symmetries 
of nonrelativistic field theory in $D+1$ dimensions.  To see that, it is 
useful to switch first to the radial coordinate $u=1/r$, which stays 
finite as we approach the naive infinity at $r\to\infty$, with the metric now
\be
\label{lifmetu}
ds^2=-\frac{dt^2}{u^{2z}}+\frac{d\bx^2+du^2}{u^2}.
\ee
For $z>1$, the $dt^2$ term is the most divergent one as we take $u\to 0$.  
In order to make the rescaled metric finite, we would like to use 
$\Omega=u^z$.  However, this choice of $\Omega$ does not have a simple zero 
at $u=0$ in this coordinate system.  In order to fix this, we change 
coordinates once again, to $w=u^z$.  The metric becomes
\be
ds^2=-\frac{dt^2}{w^2}+\frac{d\bx^2}{w^{2/z}}+\frac{dw^2}{z^2w^2}.
\ee
We can now use $\Omega=w$ to rescale the metric, but the resulting geometry 
\be
\tilde{ds}^2=-dt^2+\frac{1}{z^2}dw^2+w^{2(1-1/z)}d\bx^2
\ee
is degenerate at the purported infinity $w=0$ when $z\neq 1$.  As a 
consequence of this rather pathological behavior of the standard notion of 
conformal infinity of the Lifshitz spacetime, it is a priori unclear how to 
perform holographic renormalization, and even how to define precisely what 
we mean by black holes in the bulk.  

This tension has been remedied \cite{aci}, for spacetimes carrying the 
additional structure of an asymptotic foliation, by generalizing Penrose's 
notion of conformal infinity to reflect the asymptotic anisotropy permitted 
by the foliation.  The basic idea is simple:  When $\CM$ carries a preferred 
foliation at least near infinity, we can use the anisotropic Weyl 
transformation (\ref{aweyl}), instead of the relativistic rescaling 
(\ref{eqnom}), to map $\CM$ inside a larger manifold $\tilde\CM$ such that 
$\bar\CM\subset\tilde\CM$.   Even in this case, the rescaling factor 
$\Omega=e^\omega$  must satisfy regularity conditions at 
$\bar\CM\setminus\CM$.  In particular, $\Omega$ must have a simple zero 
there.  With a judiciously chosen value of $z$, the {\it anisotropic 
conformal infinity} $\bar\CM\setminus\CM$ can be of codimension one.  
Moreover, it naturally inherits a preferred ``anisotropic conformal 
structure,'' with conformal transformations given by those 
foliation-preserving diffeomorphisms that preserve the boundary metric up 
to an anisotropic Weyl rescaling.

Both Lifshitz and Schr\"odinger spacetimes belong to this 
class of asymptotically foliated geometries, and the resulting notion 
of anisotropic conformal infinity matches the intuitive expectations from 
holography \cite{aci}.  In the case of the Lifshitz spacetime (\ref{lifmet}), 
we start again with the metric as given in (\ref{lifmetu}).  
We interpret this geometry as carrying a natural codimension-one foliation by 
leaves of constant $t$, at least near $u\to 0$.  As we saw in 
Section~\ref{awt}, this additional structure of an asymptotic foliation gives 
us the additional freedom to use anisotropic Weyl transformations \ref{aweyl} 
without violating the symmetries.  Choosing the rescaling factor
\be
\Omega=u
\ee
and applying the anisotropic Weyl transformation (\ref{aweyl}) maps the 
Lifshitz metric in the asymptotic regime of $u\to 0$ to the flat metric, 
\be
\label{flmet}
\tilde{ds}^2=-dt^2+d\bx^2+du^2.
\ee
$u$ can now be analytically extended from $u>0$ to all real values.  
The anisotropic conformal infinity of the $(D+2)$-dimensional Lifshitz 
spacetime is at $u=0$.  Topologically, it is ${\bf R}^{D+1}$, and very 
similar to the conformal infinity of the Poincar\'e patch of $AdS_{D+2}$.  
However, even though the induced metric on anisotropic conformal infinity at 
$u=0$ in (\ref{flmet}) 
looks naively relativistic, one must remember that its natural symmetries are 
not relativistic: This conformal infinity carries a preferred foliation by 
leaves of constant $t$, and a natural anisotropic conformal structure 
characterized by dynamical exponent $z$.  The natural symmetries are given by 
those foliation-preserving diffeomorphisms that preserve the metric up to 
an anisotropic Weyl transformation \cite{aci}.  In addition to the spatial 
rotations and spacetime translations of ${\bf R}^{D+1}$, one can easily 
check that this symmetry group contains also the anisotropic scaling 
transformations (\ref{ascaling}).  Thus, the conformal structure of 
anisotropic conformal infinity nicely matches the expected conformal 
symmetries of the dual field theory.  

\subsection{Asymptotically Lifshitz Spacetimes}

Equipped with the notion of anisotropic conformal infinity of spacetime, 
we can now give a precise definition of spacetime geometries that are 
``asymptotically Lifshitz.''  Simply put, given a value of $z$, a spacetime is 
said to be asymptotically Lifshitz if it exhibits the same anisotropic 
conformal infinity as the Lifshitz spacetime for that value of dynamical 
exponent $z$.  This definition follows the logic that leads to the notions of 
asymptotic flatness and asymptotic $AdS$ \cite{hawkinge,penrin}, and extends 
such notions naturally to the case of anisotropic scaling.  

As a part of their definition, the spacetimes which are asymptotically 
Lifshitz must carry an asymptotic foliation structure near their anisotropic 
conformal infinity.  In the context of holographic renormalization, this 
condition translates into an important restriction on the form of the vielbein 
fall-off, 
\be
\label{folcd}
\frac{e_i^0}{r^z}\to 0 \qquad {\rm as}\ r\to\infty.
\ee
This provides an answer to a question discussed in \cite{ross}:  Our 
definition of asymptotically Lifshitz spacetimes using the notion of 
anisotropic conformal infinity requires that the sources for the 
energy flux vanish.%
\footnote{More precisely, it would be sufficient to impose $(\p_i e_j^0-
\p_j e_i^0)/r^z\to 0$ at infinity, a constraint which also emerges naturally 
in the vielbein formulation of gravity with anisotropic scaling.   In this 
paper, we impose the stronger condition (\ref{folcd}).}

With the definition of ``asymptotically Lifshitz'' at hand, it is now possible 
to define precisely black holes and their event horizons in Lifshitz 
spacetimes, by referring to the properties of the anisotropic conformal 
infinity of spacetime just as in the more traditional spacetimes which have 
codimension-one isotropic conformal infinity.  We refer the reader 
to \cite{aci} for additional results, and to Appendix~\ref{appbsfaas} for 
a summary of the asymptotic behavior of fields in the asymptotically Lifshitz 
spacetimes relevant for the rest of this paper.  

\section{Holographic Renormalization in Asymptotically Lifshitz Spacetimes}
\label{sechr}

Holographic duality in asymptotically $AdS$ spacetimes%
\footnote{For a pedagogical introduction, see the TASI lectures \cite{tasim} 
and \cite{tasip}.}
-- or, by logical extension, in asymptotically Lifshitz spacetimes -- relates 
the partition function of a bulk gravity system with Dirichlet boundary 
conditions at conformal infinity to the generating function of correlators in 
the appropriate dual quantum field theory.  At low energies and to leading 
order, this correspondence gives the connected generating functional $W$ with 
sources $f^{(0)}$ on the field theory side, in terms of the on-shell bulk 
gravity action evaluated with Dirichlet boundary conditions given by $f^{(0)}$: 
\be
\label{hduals}
W[f^{(0)}]=-S_{\rm on\mhyphen shell}[f^{(0)}].
\ee
Both sides of this correspondence are divergent:  Standard ultraviolet 
divergences appear on the field theory side, and they require conventional 
renormalization.  This behavior is matched on the gravity side, where the 
divergences are infrared effects, due to the scales that diverge as we 
approach the spacetime boundary.  Holographic renormalization 
\cite{mans,vijayper,perfinn,dhss,bfs,dbvv} (for reviews, see 
\cite{revdb,revsk,revjp}) is the technology designed to perform the subtraction 
of infinities on the gravity side, in the form of divergent boundary terms 
in the on-shell action, and to make precise sense of (\ref{hduals}).

Recent papers \cite{ross,mann,deboer} have performed various steps of 
holographic renormalization in Lifshitz spacetime at the non-linear level, 
and our paper builds on the results established there.  Since we choose for 
our analysis the Hamiltonian approach to holographic renormalization 
\cite{papask,papaske}, our treatment is closest to that of \cite{ross}.  

\subsection{Hamiltonian Approach to Holographic Renormalization}

The original analysis of holographic renormalization relied on properties of 
asymptotic expansions near the conformal infinity of spacetime 
\cite{ffg,graham,gw}.  The Hamiltonian approach of \cite{papask,papaske} 
aspires to give a somewhat more covariant picture, and the results of the 
earlier asymptotic expansion approach can be reproduced from it \cite{papask}. 
Either way, we start by choosing a radial coordinate, $r$, in some 
neighborhood of the anisotropic conformal infinity of the Lifshitz spacetime 
$\CM$, such that the hypersurfaces of constant $r$ are diffeomorphic to the 
boundary $\p\CM$, and they equip $\CM$ near $\p\CM$ with a codimension-one 
foliation structure.%
\footnote{In our conventions, $\p\CM$ is at $r=\infty$.  The choice of 
$u=1/r$ instead of $r$ as a coordinate near $\p\CM$ would be more appropriate, 
since $u$ is finite through $\p\CM$.  In this section, we leave this more 
rigorous coordinate choice implicit.}
This foliation should not be confused with the preferred folation of the 
anisotropic conformal boundary by leaves of constant $t$ -- the asymptotic 
regime of our spacetime carries a nested foliation structure, with leaves 
of constant radial coordinate $r$ further foliated by leaves of constant $t$.

Our starting point is the theory of bulk gravity in $3+1$ dimensions%
\footnote{The case of general $D$ and $z$ is discussed in 
Appendix~\ref{appnot}.} 
with negative $\Lambda$, coupled to some matter $\Phi$ to be specified later.  
The action is given by
\be
S_{\rm bulk}=\frac{1}{16\pi G_4}\int_\CM dt\,d^2\bx\,dr\,\sqrt{-G}
\left(\CR-2\Lambda\right)+\frac{1}{8\pi G_4}\int_{\p\CM} dt\,d^2\bx\,
\sqrt{-g}\, K+S_{\rm matter}[\Phi,G].
\ee
We will work throughout in the radial gauge, setting the radial lapse to $1/r$ 
and the radial shift to zero, in some neighborhood of the boundary (see 
Appendix~\ref{appnot} for a detailed summary of our notation and conventions).  

Our task is to evaluate the on-shell action as a functional of the boudary 
fields, and perform the corresponding renormalization.  
Because of the infinite volume of Lifshitz space, the on-shell action 
diverges, and must first be regularized by inserting a cutoff at finite volume 
and indentifying terms that diverge in the asymptotic expansion in the cutoff, 
and  then renormalized by and introducing appropriate counterterms to 
eliminate the divergences.  The on-shell action is regulated by cutting the 
bulk spacetime off at some 
value $r<\infty$ of the radial coordinate.  If $\CM_r$ is the cut-off manifold, 
its boundary $\p\CM_r$ represents a regulated boundary of spacetime.  The 
on-shell action is a function of the regulator $r$, and the boundary fields 
which include the metric multiplet $N,N_i,\gamma_{ij}$ plus all sources 
associated with the bulk matter $\Phi$, which we collectively denote by 
$\phi$.  From now on, we simply denote the on-shell action $\Sonshell$ -- 
viewed as a functional of the boundary values of the fields -- by $S$, and 
parametrize it as
\be
S=\frac{1}{16\pi G_4}\int dt\,d^2\bx\,\sqrt{\gamma}\,N\CL.
\ee
Since the on-shell action $S$ is a function of $r$ and the boundary 
values of the fields, we can naturally interpret it as a solution to the 
Hamilton-Jacobi equation, regarding $r$ as the evolution parameter.  This 
is the starting point for the Hamiltonian approach to holographic 
renormalization.  The Hamilton-Jacobi theory implies that the first variation 
of the on-shell action with respect to the boundary fields gives the conjugate 
momenta.  In the holographic dictionary, the boundary fields serve as sources, 
and their conjugate momenta are thus directly related to the one-point 
functions of the operators conjugate to the sources.  

A convenient way of computing the divergent part of $\CL$  is to organize the 
terms with respect to their scaling with $r$. More precisely, we define the 
dilatation operator by
\be
\deltaD=\int_{\p\CM_r}dt\,d^2\bx \left(zN\frac{\delta}{\delta N}
+ 2N_i\frac{\delta}{\delta N_i}+ 2\gamma_{ij}\frac{\delta}{\delta \gamma_{ij}} 
-\sum_\phi \Delta_\phi \phi \frac{\delta}{\delta\phi},\right),
\ee
where $\Delta_\phi$ collectively denotes the asymptotic decay exponents of the 
bulk matter fields $\Phi$.  Quantities of interest can then be decomposed into 
a sum of terms with definite scaling dimension under $\deltaD$.  For example, 
the object of our central interest, $\CL$, can be expanded as
\be
\label{llexp}
\CL=\sum_{\Delta}\CL^{(\Delta)}+\tilde\CL^{(z+2)}\log r.
\ee
Throughout this paper, superscripts in parentheses on any object $\CO$ 
always denote the scaling dimension in the decomposition of $\CO$ 
as a sum of terms of definite engineering scaling dimensions.  
For example, $T^{(0)}_{AB}$ is the constant part of the stress tensor,  
and $R^{(2)}$ is the dimension-two part of the scalar curvature.

The individual terms of the expansion (\ref{llexp}) satisfy
\be
\deltaD\CL^{(\Delta)}=-\Delta\,\CL^{(\Delta)}\qquad{\rm for}\ \ \Delta\ne z+2.
\ee
When $\Delta=z+2$, the scaling behavior is anomalous, 
\be
\deltaD\CL^{(z+2)} = -(z+2)\CL^{(z+2)} + \tilde\CL^{(z+2)},
\ee
with the inhomogeneous term satisfying 
\be
\deltaD\tilde\CL^{(z+2)} =- (z+2)\tilde\CL^{(z+2)}.
\ee
This logarithmic term in (\ref{llexp}) reflects the possibility of 
an anisotropic Weyl anomaly.  

The dynamical equations for the divergent part of $\CL$ are determined as 
follows.  Since the on-shell action satisfies the Hamilton-Jacobi equation, 
its radial derivative is determined in terms of the Hamiltonian.  
Because in our case the fields have fixed asymptotic behavior 
(see Appendix~\ref{appbsfaas}), in the asymptotic region the radial derivative 
is equivalent to the anisotropic scaling operator, 
\be
r\frac{d}{dr}\approx\deltaD.  
\ee
The Hamilton-Jacobi equation then 
relates the action of $\deltaD$ on the on-shell action to the Hamiltonian 
of the system.  In our case, with relativistic gravity in the bulk, the 
equation of motion for the radial lapse gives the Hamiltonian constraint.  
Using the bulk equations of motion, one obtains a first-order differential 
equation for $\CL$ in terms of the boundary values of the fields that can be 
solved iteratively in the expansion in eigenmodes of $\deltaD$.

Equivalently, one can expand the Hamiltonian constraint in eigenmodes of 
$\deltaD$.  The structure of these equations allows for the momentum modes 
to be obtained recursively in terms of the boundary data.  In this method, the 
dilatation operator acting on the on-shell action gives an expression linear 
in the canonical momenta, so that the values for the momenta obtained 
recursively from the Hamiltonian constraint give rise directly to the 
desired expression on-shell action.  The resulting on-shell action will have 
divergent pieces that can be expressed as local functionals of the boundary 
data.  These pieces can be subtracted, leading to the finite renormalized 
on-shell action.  

Further technical details of the procedure for determining the coefficients 
$\CL^{(\Delta)}$ and $\tilde\CL^{(z+2)}$ are summarized in 
Appendix~\ref{apphre}.  

\subsection{Bulk Gravity with a Massive Vector}
\label{bgmv}

We begin with the theory of relativistic bulk gravity in $3+1$ dimensions, 
coupled to a bulk massive vector field $\CA_\mu$.  The action is
\bea
S_{\rm bulk}&=&\frac{1}{16\pi G_4}\int_\CM dt\,d^2\bx\,dr\,\sqrt{-G}
\left(\mathcal{R}-2\Lambda-\frac{1}{4}F_{\mu\nu}F^{\mu\nu}
-\frac{1}{2}m^2\CA_\mu\CA^\mu\right)\cr
&&\qquad\qquad+\frac{1}{8\pi G_4}\int_{\partial\CM} dt\,d^2\bx\,\sqrt{-g}\,K.
\eea
As summarized in Appendix~\ref{appnot}, this theory has the Lifshitz 
spacetime as a solution.  In this theory, the boundary data we can specify 
reduce to the metric multiplet $(N,N_i,\gamma_{ij})$ -- or, alternatively, 
their vielbein counterparts (see Appendix~\ref{apadm}) -- and a scalar 
source $\psi$ for the massive vector.

Although our main interest will be in $z=2$, we start by considering general 
$z$.  
If we set $\psi=0$, the terms that will give rise to divergent contributions 
in the on-shell action for $z<4$ are $\CL^{(0)}$, $\CL^{(2)}$, $\CL^{(2z)}$, 
and $\CL^{(4)}$.  The holographic renormalization equations, found in 
in \cite{ross} (and reviewed in Appendix~\ref{apphre}), take the form
\bea
\CL^{(0)} &=& 2(z+1), \\
z\CL^{(2)} &=& R^{(2)} - \frac{1}{4} (F_{AB} F^{AB})^{(2)}, \\
(2-z)\CL^{(2z)} &=& R^{(2z)} + \frac{1}{2m^2}\left((\nabla_A\pi^A)^{(z)}
\right)^2 \\
(z-2)\CL^{(4)} &=& K^{(2)}_{AB}\pi^{AB(2)} + \frac{1}{2}\pi^{A(2)}{\pi_A}^{(2)} .
\eea 
With some effort these can be computed in terms of the boundary metric 
multiplet $(N,N_i,\gamma_{ij})$, giving (up to total derivatives)
\bea
\CL^{(0)}&=&2(z+1),\\
z\CL^{(2)}&=&\hat{R}+\frac{\alpha^2}{2}\left(\frac{\nabla N}{N}\right)^2,\\
(2-z)\CL^{(2z)}&=&\hat{K}_{ij}\hat{K}^{ij}+\frac{z-3}{2}\hat{K}^2,\\
(2-z)\CL^{(4)}&=&\frac{z-2}{2z^4(z+1)(z-2+\beta_z)^2}\left\{
-4z(z-6+\beta_z)\left(\frac{\nabla^2 N}{N}\right)^2\right.\cr
&&{}+\left(12+36z-11z^2-2z^3+5z^4+\beta_z(z^3-7z-2)\right)
\left[\vphantom{\frac{N}{N}}\right.\left(\frac{\nabla N}{N}\right)^2
\left.\vphantom{\frac{N}{N}}\right]^2\cr
&&{}-2z\left(36-4z-7z^2+5z^3+\beta_z(z^2-z-6)
\right)\frac{\nabla^2N}{N}\left(\frac{\nabla N}{N}\right)^2\cr
&&\qquad\left.{}+(z-6+\beta_z)\left[\vphantom{\frac{N}{N}}\right.
4 z^2\left(\frac{\nabla N}{N}\right)^2\hat{R}
-4 z^2\frac{\nabla^2 N}{N}\hat{R}
-z^3 \hat{R}^2\left.\vphantom{\frac{N}{N}}\right]\right\}. 
\eea
When $z=2$ is approached, the divergent terms of dimension four become 
logarithmic, and  the residue of the $\Delta=4$ (or $\Delta=2z$) terms at 
the $z=2$ pole give rise to $\tilde\CL^{(4)}$. Specifically, we get 
\bea
 \tilde\CL^{(4)} =\lim_{z\rightarrow 2}\left[(z-2)\CL^{(4)}+(2-z)\CL^{(2z)}
\right].
\eea 
With this substitution, the $z=2$ divergent terms in the on-shell action are
\bea
\CL^{(0)} &=& 2(z+1) = 6, \\
\CL^{(2)} &=& \frac{1}{2}\hat{R}+ \frac{1}{4}
\left(\frac{\nabla N}{N}\right)^2, \\
\tilde\CL^{(4)} &=& \hat K_{ij}\hat K^{ij} - \frac{1}{2}\hat K^2. 
\eea
The coefficient $\tilde\CL^{(4)}$ of the logarithmic divergence can be 
recognized as the unique kinetic term (\ref{confk}) for Lifshitz gravity with 
local conformal invariance in $2+1$ dimensions, invariant under the $z=2$ 
anisotropic Weyl transformations (\ref{aweyl}).  This is one of the central 
results of this paper.

The expression for the counterterms has no potential term -- \ie, the only 
derivatives that appear in the counterterm are the time derivatives.  
This is in spite of the fact that there exists a term with spatial 
derivatives, invariant under the local $z=2$ anisotropic Weyl transformations, 
\be
\label{spanom}
\int \! dt\, d^2\bx \,\sqrt{g}\,N \left\{\hat{R}+\frac{\nabla^2 N}{N}-
\left(\frac{\nabla N}{N}\right)^2 \right\}^2 ,
\ee
which is not a total derivative.

It is surprising, at least at first sight, that such a potential term is not 
generated in the logarithmic counterterm of holographic renormalization in 
Lifshitz space.  Indeed, as we show in Appendix~\ref{appwano}, this term 
(\ref{spanom}) represents a non-trivial cohomology class appropriate to appear 
as an anomaly.  What would be a minimal generalization of our holographic 
setup, which would generate such a term in the anomaly?  One might suspect 
that a different dynamical embedding of the Lifshitz space may perhaps 
produce a more general set of holographic counterterms, allowing 
(\ref{spanom}) to appear.  Even in the embedding considered here, we have not 
turned on the most general sources in the boundary, and one can ask whether 
allowing nonzero $\psi$ generates new counterterms.  However, a detailed 
calculation (see Appendix~\ref{apphre}) reveals that turning on $\psi$ also 
preserves detailed balance, and does not lead to the appearance of 
the second independent counterterm (\ref{spanom}).  

\subsection{Gravity with a Massive Vector Coupled to Bulk Scalars}
\label{bgmvs}

In order to probe further the structure of holographic counterterms in 
Lifshitz spacetime, it is useful to add additional matter fields in the bulk 
theory.  The holographic renormalization procedure can be easily repeated with 
the inclusion of scalar fields in the bulk.  We will see that for a marginal 
scalar at $z=2$, there is a new logarithmically divergent counterterm, giving 
rise to a new, nongravitational contribution to the anisotropic Weyl anomaly.  
However, we will see that this new counterterm also satisfies the detailed 
balance condition:  Even in the presence of the bulk scalars, the second 
gravitational counterterm (\ref{spanom}) -- which violates detailed balance 
-- is not generated.  

The bulk scalar action takes the standard relativistic form
\be
S_{{\rm bulk,}\ X}=-\frac{1}{2}\int_{\CM} d^{d+1}x \sqrt{-G} \left(G^{\mu\nu}
\partial_\mu X^a\partial_\nu X^a+\mu^2 X^aX^a\right).
\ee
In this section, we set $d=3$, and again follow the procedure of \cite{ross}, 
with appropriate modifications to include the scalar fields.  The holographic 
renormalization equations of \cite{ross} now become
\be
(z+2-\Delta)\CL^{(\Delta)}= \tilde{\FQ}^{(\Delta)}+\tilde{\FS}^{(\Delta)},
\ee
where the quadratic and source terms $\FQ$ and $\FS$ are modified to 
\be
\tilde{\FQ}^{(\Delta)}=\FQ^{(\Delta)}+8\pi G_4
(\tilde{\pi}^{a(\Delta/2)})^2+16\pi G_4\sum_{s<\Delta/2;\ s\neq
\tilde{\Delta}_-}
(\tilde{\pi}^{a(s)}\tilde{\pi}^{a(\Delta-s)})
\ee
and
\be
\tilde{\FS}=\FS-8\pi G_4(\partial_\alpha X^a\partial^\alpha X^a
+\mu^2 X^aX^a).
\ee
In this expression, $\tilde{\pi}^{a}=r\partial_r X^a$ is the scalar 
momentum and the scalars fall of asymptotically as $r^{-\tilde{\Delta}_-}$, 
where 
$$\mu^2=\tilde{\Delta}_-(\tilde{\Delta}_--2-z).$$

The additional source terms only contribute at orders 
$\Delta=2\tilde{\Delta}_-,$ $2+2\tilde{\Delta}_-$ and $2z+2\tilde{\Delta}_-$:
\bea
\tilde{\FS}^{(2\tilde{\Delta}_-)}&=&-8\pi G_4\mu^2 X^aX^a,\\
\tilde{\FS}^{(2+2\tilde{\Delta}_-)}&=&-\left[8\pi G_4\partial_\alpha 
X^a\partial^\alpha X^a\right]^{(2+2\tilde{\Delta}_-)}=-8\pi G_4
\partial_i X^a\partial^i X^a,\\
\tilde{\FS}^{(2z+2\tilde{\Delta}_-)}&=&
-\left[8\pi G_4\partial_\alpha X^a\partial^\alpha X^a
\right]^{(2z+2\tilde{\Delta}_-)}=\frac{8\pi G_4}{N^2}
(\partial_t X^a-N^i\partial_i X^a)^2.
\eea
We now specialize to the case of a marginal scalar, that is, a scalar which 
has $\tilde{\Delta}_-=0$. Note that this also means that the scalar is 
massless  
since $\mu^2=\tilde{\Delta}_-(\tilde{\Delta}_--2-z)=0$. We are interested in 
calculating the contribution to the anisotropic Weyl anomaly in the case 
$z=2$. The divergent pieces of the on-shell action that appear at orders 
$\Delta=2+2\tilde{\Delta}_-$ and $\Delta=2z+2\tilde{\Delta}_-$ are 
straightforward to calculate as they only receive contributions from the 
source terms,
\bea
(z-2\tilde{\Delta}_-)\CL^{(2+2\tilde{\Delta}_-)}&=&-8\pi G_4
\partial_i X^a\partial^i X^a,\\
(2-z-2\tilde{\Delta}_-)\CL^{(2z+2\tilde{\Delta}_-)}&=&
\frac{8\pi G_4}{N^2}(\partial_t X^a-N^i\partial_i X^a)^2.
\eea
By taking the functional derivative of this term in the on-shell action with 
respect to the metric, the contribution to the boundary stress energy tensor 
can be calculated. For example, for $\Delta=2+2\tilde{\Delta}_-$
\bea
(z-2\tilde{\Delta}_-)T_{00}^{(2+2\tilde{\Delta}_-)}&=&-8\pi G_4
\partial_i X^a\partial^i X^a,\\
(z-2\tilde{\Delta}_-)T_{IJ}^{(2+2\tilde{\Delta}_-)}&=& -16\pi G_4\partial_I X^a
\partial_J X^a+ 8\pi G_4\partial_i X^a\partial^i X^a\delta_{IJ},\\
(z-2\tilde{\Delta}_-)T_{0I}^{(3-z+2\tilde{\Delta}_-)}&=& 0.
\eea
In addition, by taking the functional derivative with respect to the scalar, 
the boundary scalar momentum can be calculated, via
$$\tilde{\pi}^a=-\frac{1}{N\sqrt{\gamma}}\frac{\delta S}{\delta X^a}.$$
For example, one gets
\bea
(z-2\tilde{\Delta}_-)\tilde{\pi}^{a(2+\tilde{\Delta}_-)}=- \frac{1}{N}
\nabla^i(N \nabla_i X^a)=-\nabla^2 X^a-\frac{\nabla^iN
\nabla_i X^a}{N}.
\eea
The higher order counterterms are more involved because they receive 
contributions from the quadratic piece. For example,
\bea
&&(z-\Delta_--2\tilde{\Delta}_-)\CL^{(2+\Delta_-+2\tilde{\Delta}_-)}
=K_{AB}^{(\Delta_-)}T^{AB(2+2\tilde{\Delta}_-)}\cr
&&\qquad\qquad=-\frac{\alpha\psi}{2(z+1)}\left[z(3z-\Delta_-)
T^{00(2+2\tilde{\Delta}_-)}+z(2z-1-\Delta_-){T^I_I}^{(2+2\tilde{\Delta}_-)}\right]\cr
&&\qquad\qquad=-\frac{\alpha\psi}{2(z+1)}\left[z(3z-\Delta_-)
T^{00(2+2\tilde{\Delta}_-)}\right],
\eea
using the fact that ${T^I_I}^{(2+2\tilde{\Delta}_-)}=0$, as calculated above. 
Note that for $z=2$ this becomes 
$\CL^{(2+\Delta_-+2\tilde{\Delta}_-)}=-\psi T^{00(2+2\tilde{\Delta}_-)}$.
The calculation of this term is useful even when the source for the massive 
vector $\psi$ is set to zero. This is because we can determine 
$\pi_\psi^{(2+2\tilde{\Delta}_-)}$ by taking the functional derivative with 
respect to $\psi$:
\be
(z-\Delta_--2\tilde{\Delta}_-)\pi_\psi^{(2+2\tilde{\Delta}_-)}=
\frac{\alpha}{2(z+1)}\left[z(3z-\Delta_-)T^{00(2+2\tilde{\Delta}_-)}\right].
\ee
The following terms also receive contributions from the quadratic piece:
\bea
(z-2-2\tilde{\Delta}_-)\CL^{(4+2\tilde{\Delta}_-)}&=&2K_{AB}^{(2)}
T^{AB(2+2\tilde{\Delta}_-)}+\pi_A^{(2)}\pi^{A(2+2\tilde{\Delta}_-)}
+8\pi G_4\left(\tilde{\pi}^{a(2+\tilde{\Delta}_-)}\right)^2,\qquad\\
(z-2-4\tilde{\Delta}_-)\CL^{(4+4\tilde{\Delta}_-)}&=&
K_{AB}^{(2+2\tilde{\Delta}_-)}T^{AB(2+2\tilde{\Delta}_-)}+\frac{1}{2}
\pi_A^{(2+2\tilde{\Delta}_-)}\pi^{A(2+2\tilde{\Delta}_-)}.
\eea
These are the terms that will contribute to the scaling anomaly when $z=2$. 
After a lengthy calculation of the right hand sides for $z=2$, the following 
result is obtained (up to total derivatives):
\bea
(z-2-2\tilde{\Delta}_-)\CL^{(4+2\tilde{\Delta}_-)}&=&2\pi G_4
(\Delta X^a)^2,\\
(z-2-4\tilde{\Delta}_-)\CL^{(4+4\tilde{\Delta}_-)}&=&\frac{1}{4}
T_{IJ}^{(2+2\tilde{\Delta}_-)}T^{IJ(2+2\tilde{\Delta}_-)},\cr
&=&16\pi^2 G_4^2\left(\partial_i X^a\partial_j X^a\partial^iX^b
\partial^j X^b- \frac{1}{2}(\partial_i X^a\partial^i X^a)^2\right).
\eea
By combining all these results, the contribution of the massless scalars to 
the logarithmically divergent counterterm when $z=2$ is (by equation 
(\ref{anomres})):
\bea
\tilde\CL^{(4)}_X&=&\lim_{z\rightarrow 2}\left((2-z)
\CL^{(2z+2\tilde{\Delta}_-)}+(z-2)
\CL^{(4+2\tilde{\Delta}_-)}+(z-2)\CL^{(4+4\tilde{\Delta}_-)}\right)\cr
&=&\frac{8\pi G_4}{N^2}(\partial_t X^a-N^i\partial_i X^a)^2
+2\pi G_4(\nabla^2 X^a)^2\cr
&&\qquad{}+16\pi^2 G_4^2\left(\partial_i X^a\partial_j X^a\partial^i
X^b\partial^j X^b- \frac{1}{2}(\partial_i X^a\partial^i X^a)^2\right).
\eea
Together with the gravitational counterterms from the previous section, 
the total counterterm action for $z=2$ is given by
\bea
S_{ct}&=&-\int_{\partial \CM}dt\,d^2\bx\sqrt{\gamma}N\left\{\frac{1}{16\pi G_4}
\left[\vphantom{\frac{N}{N}}\right.
6+\frac{1}{2}\hat{R}+\frac{1}{4}\left(\frac{\nabla N}{N}\right)^2
\left.\vphantom{\frac{N}{N}}\right]
-\frac{1}{4}\partial_i X^a\partial^i X^a\right.\cr
&&\qquad{}-\log\epsilon\left[\frac{1}{16\pi G_4}(\hat{K}_{ij}\hat{K}^{ij}-\frac{1}{2}\hat{K}^2)+
\frac{1}{2N^2}(\partial_t X^a-N^i\partial_i X^a)^2+\frac{1}{8}
(\nabla^2 X^a)^2\right.\cr
&&\qquad\qquad{}+\left.\left.\pi G_4\left(\partial_i X^a
\partial_j X^a\partial^iX^b\partial^j X^b- \frac{1}{2}(\partial_i 
X^a\partial^i X^a)^2\right)\right]\right\}.
\eea
Interestingly, this logarithmically divergent counterterm takes the form 
identical to the action written down in \cite{mqc}, describing the coupling 
of $z=2$ gravity and $z=2$ Lifshitz matter in $2+1$ dimensions.  This action 
is invariant under $z=2$ anisotropic Weyl transformations, with the scalars 
transforming with weight zero, and satisfies the detailed balance condition.  
As a result, it was shown in \cite{mqc} that the ground-state wavefunction of 
this membrane action on a spatial surface $\Sigma$ is given by the bosonic 
string partition function on $\Sigma$.  We see that the property of detailed 
balance, satisfied by the logarithmic counterterms in the absence of extra 
matter, persists in the presence of the marginal scalar fields.  

Two additional comments are worth making: 

(1) The relative sign between the potential terms and the kinetic term in the 
logarithmic counterterm is opposite to the sign one would expect from the 
action of a unitary theory with $z=2$ scaling in real time.  This is not very 
surprising, and corresponds to the 
fact already appreciated in the relativistic case:  The holographic 
counterterms do not have to reproduce the action of a unitary theory, as is 
clear from the appearance of the higher-derivative conformal gravity action 
in the holographic counterterms in $AdS_5$.

(2) In the classical theories with Lifshitz scaling, the coupling 
constants in front of the individual contributions to the potential term 
are not related by any symmetry to the kinetic terms.  Therefore, they 
represent classically marginal couplings.  In the structure of our 
counterterms, we find this freedom realized only partially:  A uniform 
overall rescaling of all the couplings in the potential can be accomplished 
by a shift in $r$, but it appears that the interaction with the bulk 
relativistic system eliminates the apparent freedom of the relative rescaling 
between different contributions to the potential from species unrelated by 
any symmetry in the boundary theory.  This mechanism deserves further study.   

\subsection{Explaining Detailed Balance}

Now that we have accumulated some evidence suggesting that the appearance of 
the detailed balance condition in the structure of the counterterms is rather 
generic, it would be desirable to obtain a more systematic explanation of 
this fact.  It would be interesting to see why this principle should be 
naturally satisfied in the context of holographic renormalization.  

A closer look at the structure of the holographic renormalization equations 
(summarized in Appendix~\ref{apphre}) reveals a simple answer:  
In the procedure we 
followed in $3+1$ bulk dimensions, the potential terms in the counterterm 
at order four are generated by quadratic terms in the stress-energy tensor and 
field momenta at order two. These momenta arise from the functional 
differentiation of the counterterm at order two. Consider the counterterm 
appearing above at order two:
\be
S_{ct}^{(2)}=-\int_{\partial\CM}dt\,d^2\bx\,\sqrt{\gamma}
N\left\{\frac{1}{32\pi G_4}\left[\vphantom{\frac{N}{N}}\right.
\hat{R}+\frac{1}{2}\left(\frac{\nabla N}{N}\right)^2
\left.\vphantom{\frac{N}{N}}\right]-\frac{1}{4}\partial_i X^a\partial^i 
X^a\right\}.
\ee
This Lagrangian is exactly the one used in the detailed balance condition in 
\cite{mqc}, in the case where $N$ does not depend upon spatial 
coordinates.%
\footnote{Detailed balance in the nonprojectable theory has been discussed 
recently in \cite{sotdb}.}
Hence, the detailed balance relation, as reviewed in Section~\ref{lifgrav}, 
is simply a consequence of the relationship between two counterterms implied 
by the holographic renormalization in asymptotically Lifshitz spacetime.  

It should be noted that in the above procedure, the presence of the massive 
vector complicates the equations 
and make the detailed-balance-like relation between the two actions less 
transparent. But the logarithmic counterterm potential terms (with scaling 
dimension four) are nonetheless directly derivable from the counterterms with 
scaling dimension two.  
  
In fact, an analogous result also holds in the 
relativistic case of holographic renormalization in $AdS_5$, where the second 
order counterterm is simply the Einstein-Hilbert action and the conformal 
anomaly is the action $S_{\rm conf}$ of conformal gravity in $3+1$ dimensions:  
It turns out that $S_{\rm conf}$ is obtained by squaring the functional 
derivative of the Einstein-Hilbert action.  The reason behind this 
relationship is the same:  $S_{\rm conf}$ and the Einstein-Hilbert action 
appear as two counterterms, linked via the holographic renormalization 
procedure into a condition reminiscent of detailed balance.

A closer look also reveals that the holographic justification for the 
detailed balance condition being satisfied by the logarithmic conterterm 
quickly ceases to be valid with increasing spacetime dimension.  However, 
this property does not disappear completely:  Instead, the holographic 
renormalization machinery implies a more complex relation between the 
logarithmic counterterm and the variational derivatives of the entire 
hierarchy of the power-law counterterms.  

\subsection{Analytic Continuation to the de~Sitter-like Regime}

In relativistic AdS/CFT correspondence, the Hamilton-Jacobi formulation of 
holographic renormalization -- with the radial direction $r$ as the evolution 
parameter -- can be easily continued analytically to de~Sitter space.  Upon 
this continuation, the evolution parameter $r$ becomes the real time $\eta$, 
and the analytic continuation of the counterterms gives useful information 
about the wavefunction $\Psi$ of the Universe on superhorizon scales 
\cite{maldai,harlow,malda}.  In particular, in the case of $AdS_5$ continued 
analytically to $dS_5$, the exponential of the logarithmic counterterm 
(known to take the form of the relativistic conformal gravity action 
$S_{\rm conf}$ in $3+1$ dimensions) is related to the wavefunction via
\be
|\Psi|^2=e^{-S_{\rm conf}}.
\ee
In this paper, we have analyzed holographic counterterms in the Lifshitz 
space background, and in the case of $z=2$ and $3+1$ bulk dimensions, we also 
found a logarithmic counterterm in the form of a $z=2$ multicritical conformal 
gravity action.  It is natural to ask whether an analytic continuation exists, 
similar to the one studied in \cite{maldai,harlow,malda}, so that the $z=2$ 
anisotropic conformal gravity action similarly produces the square of the 
wavefunction of the dual system.  The answer appears to be yes, and the dual 
system is a gravity theory with an interesting kind of spatial anisotropy.  

Reintroducing the length scale $L_r$ in the spacetime metric of the Lifshitz 
space at $z=2$, 
\be
ds^2=L_r^2\left(-r^{4}dt^2+r^2d\bx^2+\frac{dr^2}{r^2}\right),
\ee
we can analytically continue our results by taking $r=i\eta$ and 
$L_r=-iL_\eta$ and relabeling $t=y$, which leads to the following spacetime: 
\be
ds^2=L_\eta^2\left(\eta^{4}dy^2+\eta^2d\bx^2-\frac{d\eta^2}{\eta^2}\right).
\ee
This spacetime can be viewed as a spatially anisotropic, ``multicritical'' 
version of de~Sitter space. We found the on-shell action for asymptotically 
Lifshitz space to be (with the cutoff at $r=1/\epsilon_r$):
\bea
S&=&\frac{L_r^2}{16\pi G_{4}}\int_{\partial\CM_{1/\epsilon_r}} 
dt\,d^2\bx\,\sqrt{\gamma}\,N(\CL^{(0)}+\CL^{(2)}+\CL^{(4)}-
\tilde{\CL}^{(4)}\log\epsilon_r)\\
&=&\frac{L_r^2}{16\pi G_{4}}\int_{\partial\CM_{1/\epsilon_r}} dt\,d^2\bx\,
\sqrt{\gamma_\fin}N_\fin\left\{\frac{\CL_\fin^{(0)}}{\epsilon_r^4}
+\frac{\CL_\fin^{(2)}}{\epsilon_r^2}+\CL_\fin^{(4)}-\tilde{\CL}_\fin^{(4)}
\log\epsilon_r\right\},
\eea
where the quantities with fins are defined to be finite as 
$r\rightarrow\infty$ (that is, $\CO^{(\D)}=\CO_\fin^{(\D)}\epsilon_r^\D$). 
The analytic continuation implies that 
the cutoff changes to $\epsilon_r=-i\epsilon_\eta$, where $\epsilon_\eta<0$. 
Note that all terms in the on-shell action remain real after the analytic 
continuation, except for the logarithm, which now has an imaginary part since 
$\log \epsilon_r=\log(-\epsilon_\eta)+i\pi/2$. Thus, after this analytic 
continuation, the square of the ground-state wavefunction for the spatially 
anisotropic version of de Sitter space is given solely by the coefficient of 
the logarithmic counterterm,

\bea
\label{gsstvw}
|\Psi|^2=|e^{iS}|^2=\exp\left\{-\frac{L_\eta^2}{16 G_{4}}\int_{\partial\CM} 
d^2\bx\,dy\,\sqrt{\gamma}\,N\,\tilde{\CL}^{(4)}\right\}.
\eea
In the case of the theory studied in Section~\ref{bgmv}, we found that 
$\tilde\CL^{(4)}$ is the action of $z=2$ conformal Lifshitz gravity 
in detailed balance.  It depends only on the $y$ derivatives but not the 
$\bx$ derivatives of the metric.  Thus, the ground-state wavefunction 
(\ref{gsstvw}) represents a theory with spatial anisotropy, ultralocal 
along all but one spatial dimension, similar to the theory discussed in 
\cite{phroma,ghmt}.  

In the theory with bulk scalars studied in Section~\ref{bgmvs}, 
$\tilde\CL^{(4)}$ was found to be the action of $z=2$ conformal Lifshitz 
gravity coupled to $z=2$ scalars, still satisfying the detailed balance 
condition.  This action has a nontrivial potential term, of fourth order in 
the $\bx$ derivatives of the scalars.  Notably, the sign of this potential 
term, which we commented on at the end of Section~\ref{bgmvs}, is such that 
the analytically continued $\tilde\CL^{(4)}$ appearing in (\ref{gsstvw}) is 
positive definite.  

\section{Conclusions}

The theory of gravity with anisotropic scaling introduced in \cite{mqc,lif} 
has already been found to play a variety of roles in condensed matter.  
For example, linearized multicritical gravity with $z=2$ and $z=3$ emerges in 
the infrared regime of various bosonic lattice models, on a rigid lattice 
\cite{eme}: Gravitons with the nonrelativistic dispersion relation represent 
low-energy collective excitations of the lattice degrees of freedom.  
Dynamical gravity with anisotropic scaling also emerges naturally from 
fermionic condensed matter systems when the fundamental fermions are 
integrated out \cite{grisha}.  In the present paper, we have added another 
role to this list:  Multicritical gravity naturally appears in the process 
of holographic renormalization of relativistic systems in spacetimes which 
are asymptotically anisotropic and describe holographic duals of 
nonrelativistic field theories.  In the process, for the special case of 
bulk $2+1$ dimensions with $z=2$, we found that holographic renormalization 
imposes the condition of detailed balance on the action of $z=2$ conformal 
gravity coupled to matter, and gives a new rationale for this -- otherwise 
somewhat obscure -- condition.    

Clearly, various interesting open questions remain.  First of all, our 
analysis of holographic renormalization in the simplest anisotropic example, 
of $z=2$ in $3+1$ bulk dimensions, should generalize straightforwardly to 
higher integer values of $z$.  Some calculations relevant for this task are 
reported in Appendix~\ref{apphre}.  In particular, at $z=3$ in $4+1$ bulk 
dimensions, we expect the appearance of logarithmic counterterms taking the 
form of the action for $z=3$ multicritical conformal gravity in $3+1$ 
dimensions, introduced in \cite{lif}.  Moreover, now that we have seen that 
the classical action of multicritical gravity appears from 
string-inspired holography, it would also be interesting to see whether the 
full {\it dynamics\/} of multicritical gravity can also be engineered from 
string theory, perhaps by taking judicious scaling limits of backgrounds 
without Lorentz invariance.  Finally, it would also be natural to extend the 
study of nonrelativistic holography to the more general case, in which the 
bulk gravity itself exhibits spacetime anisotropies and multicriticality.%
\footnote{Some early steps in that direction were suggested in \cite{mqc,lif}.}
Such constructions could extend 
the list of nonrelativistic field theories amenable to a holographic 
description to a broader class, in which those nonrelativistic theories that 
have a relativistic bulk dual may well be only a minority.

\acknowledgments
We wish to thank Yu Nakayama and Omid Saremi for useful discussions.  
This work has been supported by NSF Grant PHY-0855653, by DOE Grant 
DE-AC02-05CH11231, and by the Berkeley Center for Theoretical Physics.  
%%%%%%%%%%%%%%%%%%%%%%%%%%%%%%%%%%%%%%%%%%%%%%%%%%%%%%%%%%%%%%%%%%%%%%

\appendix

\section{Notation and Conventions}
\label{appnot}

We use the following bulk metric:
\bea
\label{metcnv}
ds^2&=&G_{\mu\nu}dx^{\mu}dx^{\nu}=g_{\alpha\beta}dx^{\alpha}dx^{\beta}
+\frac{dr^2}{r^2}\cr
&=&-N^2 dt^2+\gamma_{ij}(dx^i+N^i dt)(dx^j+N^j dt)+\frac{dr^2}{r^2}.
\eea
The boundary is at $r=\infty$. $D$ is the number of spatial dimensions on the 
boundary and so there are $D+2$ spacetime dimensions in the bulk and $d\equiv 
D+1$ spacetime dimensions on the boundary. For coordinate indices, $i,j$ are 
used for the $D$ spatial boundary indices ($x^i$), whereas $\alpha,\beta$ are 
used for the $D+1$ spacetime boundary indices ($t,x^i$) and $\mu,\nu$ are 
used for the $D+2$ bulk dimensions ($t,x^i,r$).   Note that in (\ref{metcnv}), 
the bulk diffeomorphisms have been gauge fixed by setting the 
the bulk shift vector $\CN_\alpha$ (defined as $\CN_\alpha=G_{r\alpha}$) to 
$\CN_\alpha=0$, and the bulk lapse function (defined via $G_{rr}=\CN^2+
g^{\alpha\beta}\CN_\alpha\CN_\beta$) to $\CN=1/r$.  This {\it radial gauge\/} 
is adopted throughout the paper.  Moreover, in order to distinguish the 
lapse and shift variables in the bulk from those of the ADM decomposition 
on the boundary, we refer to the bulk variables $\CN$ and $\CN_i$ as the 
``radial lapse'' and ``radial shift.''  

It is often convenient to work in terms of vielbeins, which we define via
\bea
ds^2&=&\eta_{MN}E_\mu^M  E_\nu^N dx^{\mu}dx^{\nu}
=\eta_{AB}e_\alpha^A e_\beta^B dx^{\alpha}dx^{\beta}+\frac{dr^2}{r^2}\cr
&=&-N^2 dt^2+\delta_{IJ}\hat{e}_i^I \hat{e}_j^J(dx^i+N^i dt)(dx^j+N^j dt)
+\frac{dr^2}{r^2}.
\eea
For the internal frame indices, $M,N=0,1,...,D+1$ are used for the $D+2$ bulk 
dimensions, $A,B=0,1,...,D$ are used for the $D+1$ spacetime boundary indices 
and $I,J=1,...,D$ are used for the $D$ spatial boundary indices.
The vielbeins allow coordinate indices to be changed to frame indices and 
vice versa, for example $F^{AB}=e_\alpha^A e_\beta^B F^{\alpha\beta}$. Also note that the vielbeins are related to the extrinsic curvature by $K_{\alpha\beta}=r(e_\alpha^A\partial_r{e}_{A\beta}+e_\beta^A\partial_r{e}_{A\alpha})/2$.

In order to distinguish the Riemann tensor and the extrinsic curvature tensor 
of the three different metrics $G_{\mu\nu}$, $g_{\alpha\beta}$ and $\gamma_{ij}$, 
we use the notation wherein $(D+2)$-dimensional quantities are written in 
curly letters (for example, $\CR$ for the Ricci scalar), $(D+1)$-dimensional 
quantities are written in standard italics and $D$-dimensional quantities are 
written with hats.

\subsection{The bulk action}

The bulk spacetime relativistic action is:
\bea
\label{baction}
S_{\rm bulk}&=&\frac{1}{16\pi G_{D+2}}\int_\CM dt\,d^D\bx\,dr\,\sqrt{-G}
\left(\mathcal{R}-2\Lambda-\frac{1}{4}F_{\mu\nu}F^{\mu\nu}
-\frac{1}{2}m^2\CA_\mu\CA^\mu\right)\cr
&&\qquad\qquad+\frac{1}{8\pi G_{D+2}}\int_{\partial\CM} dt\,d^D\bx\,\sqrt{-g}
\,K.
\eea
Note that in order for the Lifshitz spacetime (\ref{lifmet}) to be a classical 
solution, we set 
\be
m^2=Dz\quad {\rm and}\quad \displaystyle{\Lambda=-\frac{1}{2}\left(z^2+(D-1)z
+D^2\right)}.
\ee
The Lifshitz metric is sourced by a non-zero condensate of the vector field, 
and we denote by $\psi$ the deviation away from this non-zero background: 
\be
\CA_A=(\alpha+\psi)\delta_A^0,
\ee
with
\be
\alpha^2=\frac{2(z-1)}{z}.
\ee
The leading order behavior of the vector at the boundary 
is  $\psi \sim r^{-\D_-}$, where we use the notation of \cite{omid,bruths}:
\be
\D_- = \frac{1}{2}(z + D - \beta_z)
\ee
and 
\be
\beta_z = \sqrt{(z + D)^2 + 8 (z - 1)(z - D)}.
\ee
When the action (\ref{baction}) is evaluated as a function of the boundary 
fields we write it as:
\bea
\label{onshact}
S=\frac{1}{16\pi G_{D+2}}\int_{\partial\CM} dt\,d^D\bx\,\sqrt{\gamma}\,N\CL.
\eea

\subsection{ADM decomposition in the metric formalism}
\label{apadma}

In our calculations, we decompose the metric $g_{\alpha\beta}$ on the 
$(D+1)$-dimensional boundary of spacetime into the ADM decomposition  
\begin{eqnarray*}
 g_{tt}=-N^2+N^iN_i, \qquad g_{ij}=\gamma_{ij}, \qquad g_{ti}= N_i, \\
 \qquad g^{tt}=-\frac{1}{N^2}, \qquad g^{ij}=\gamma^{ij}
-\frac{N^iN^j}{N^2}, \qquad g^{ti}=\frac{N^i}{N^2}.
\end{eqnarray*}

This metric leads to the following Christoffel symbols:
\begin{eqnarray*}
 \Gamma_{tt}^t &=&\frac{\partial_tN}{N} +\frac{N^j\nabla_jN}{N}+\frac{N^iN^j
\hat{K}_{ij}}{N},\\
  \Gamma_{tt}^i  &=&\gamma^{ij}N\nabla_jN +N\gamma^{ij}
\partial_t\left(\frac{N_j}{N}\right
)-\frac{N^iN^j\nabla_jN}{N}-\gamma^{ij}N^k\nabla_jN_k
-\frac{N^iN^jN^k\hat{K}_{jk}}{N},\\
\Gamma_{ti}^t&=&\frac{\nabla_iN}{N}+\frac{N^j\hat{K}_{ij}}{N},\\
 \Gamma_{ti}^j &=&N\gamma^{jk}\hat{K}_{ik}+N\nabla_i\left(\frac{N^j}{N}\right)
-\frac{N^jN^k\hat{K}_{ik}}{N},\\
 \Gamma_{ij}^t &=&\frac{\hat{K}_{ij}}{N}, \\
 \Gamma_{ij}^k &=&\hat{\Gamma}_{ij}^k-\frac{\hat{K}_{ij}N^k}{N},
 \end{eqnarray*}
 where $\displaystyle \hat{K}_{ij}=\frac{1}{2N}(\partial_t\gamma_{ij}
-\nabla_i N_j-\nabla_j N_i)$ is the $D$-dimensional extrinsic curvature.

These result in the following $(D+1)$-dimensional Ricci scalar $R$ for the 
metric $g_{\alpha\beta}$ in terms of $\hat{R}$, the $D$-dimensional Ricci 
scalar for the metric $\gamma_{ij}$:
\be
\label{riccidec}
R=\hat{R}-\frac{2\nabla^2N}{N}+\hat{K}_{ij}
\hat{K}^{ij}-\hat{K}^2 +\frac{\partial_tZ}{N\sqrt{\gamma}}
+\frac{\nabla^iY_i}{N},
\ee
where:
\bea
Z&\equiv&\gamma^{ij}\sqrt{\gamma}\nabla_i\left(\frac{N_j}{N}\right)
+2\hat{K}\sqrt{\gamma},\\
Y_i&\equiv&-\partial_t\left(\frac{N_i}{N}\right)+\frac{N^j\nabla_iN_j}{N}
+2N^j\hat{K}_{ij}
-3N_i\hat{K}+\frac{N^j\nabla_jN_i}{N}-\frac{N_i\nabla_jN^j}{N}.
\eea

\subsection{ADM decomposition in the vielbein formalism}
\label{apadm}

The vielbeins are defined by $g_{\alpha\beta}=e^A_\alpha e^B_\beta\eta_{AB}$ 
and $\gamma_{ij}=\hat{e}^I_i\hat{e}^J_j\delta_{IJ}$. The $(D+1)$ dimensional 
boundary has vielbeins $e^A$ given by:
\bea
e^{0}=N dt, \qquad e^{I}=\hat{e}^{I}_i(N^i dt +dx^i)=N^I dt+\hat{e}^I.
\eea

The Ricci rotation coefficients are defined by $de^C={\Omega_{AB}}^Ce^A\wedge e^B$,
\bea
de^0&=&\nabla_iN dx^i\wedge dt=\frac{\nabla_IN}{N} e^I\wedge e^0,\\
de^I&=&\left(\frac{\nabla_JN^I}{N}-\frac{\hat{e}_J^j\partial_t\hat{e}_j^I}{N}
\right) e^J\wedge e^0+{\hat{\Omega}_{JK}}^{\textrm{ \quad }I} e^J\wedge e^K.
\eea
This means that:
\bea
{\Omega_{0I}}^0&=&\frac{\nabla_IN}{2N},\\
{\Omega_{IJ}}^0&=&0,\\
{\Omega_{0J}}^I&=&-\frac{\nabla_JN^I}{2N}+\frac{\hat{e}_J^j\partial_t\hat{e}_j^I}{2N},\\
{\Omega_{JK}}^I&=&{\hat{\Omega}_{JK}}^{\textrm{ \quad }I},  \\
{\Omega_{0I}}^I&=&-\frac{\nabla_IN^I}{2N}+\frac{\hat{e}_I^j\partial_t\hat{e}_j^I}{2N}=-\frac{\nabla_IN^I}{2N}+\frac{\gamma^{ij}\partial_t{\gamma_{ij}}}{4N}=\frac{\hat{K}}{2}.
\eea
Note that by definition ${\Omega_{AB}}^C=-{\Omega_{BA}}^C$.
The covariant derivative is then given by: 
\bea
\nabla_\alpha V_B=\partial_\alpha V_B-{\omega_{\alpha B}}^CV_C,
\eea
where $\omega_{ABC}=-\Omega_{ABC}+\Omega_{ACB}+\Omega_{BCA}$. 
Note that $\omega_{ABC}=-\omega_{ACB}$. Also ${\omega_{[AB]}}^C=-{\Omega_{AB}}^C$ and ${\omega_{CD}}^C=2{\Omega_{DC}}^C$.

\subsection{The massive vector}
\label{apmv}

We take the massive vector to be $\CA_A=(\alpha+\psi)\delta_A^0$ where $\alpha^2=2(z-1)/z$. Also, the massive vector has a non-zero component in the $r$ direction, which the equation of motion for $\CA_r$ gives as $\displaystyle\CA_r=-\frac{\nabla^\alpha F_{r\alpha}}{m^2}=-\frac{\nabla^A\pi_A}{m^2r}$. Then:
\bea
\CA_\alpha&=&e_\alpha^A\CA_A=e_\alpha^0(\alpha+\psi)= N (\alpha+\psi)\delta_\alpha^t.
\eea
The only non-zero component of $F_{\alpha\beta}$ is
\bea
F_{it}=-F_{ti}=\partial_i\CA_t= \alpha\nabla_i N +\nabla_i (N\psi).  
\eea
The non-zero components of $F^{\alpha\beta}$ are
\bea
F^{jt}=-F^{tj}=-\frac{\gamma^{ij}F_{it}}{N^2}, \qquad
F^{jk}=\frac{(\gamma^{ij}N^k-\gamma^{ik}N^j)F_{it}}{N^2}.
\eea
Therefore we have that:
\bea
\label{fsquareddec}
F_{AB}F^{AB}=F_{\alpha\beta}F^{\alpha\beta}&=&-\frac{2(\alpha\nabla_i N +\nabla_i (N\psi) )(\alpha\nabla^i N +\nabla^i (N\psi) )}{N^2} \cr
&=&-2\alpha^2\left(\frac{\nabla N}{N}\right)^2-4\alpha\frac{\nabla_i (N\psi)\nabla^i N}{N^2} -\frac{2\nabla_i (N\psi)\nabla^i (N\psi) }{N^2}.
\eea

\subsection{Functional derivatives and the stress tensor}
\label{ape}
We define the momenta corresponding to the metric $g_{\alpha\beta}$ and vector field $\CA_\alpha$ by $\pi_{\alpha\beta}=K_{\alpha\beta}-g_{\alpha\beta}K$ and $\pi_\alpha=rF_{r\alpha}$ respectively,%
\footnote{This differs from the usual canonical momenta by a factor of $\sqrt{-g}(16\pi G_{D+2})^{-1}$ in order to simplify some of the subsequent equations.} 
where $K_{\alpha\beta}=r\partial_r g_{\alpha\beta}/2$.
As in the standard Hamilton-Jacobi theory, the momenta can also be obtained by functional differentiation of the on-shell action:
\bea
\pi^{\alpha\beta}=-\frac{16\pi G_{D+2}}{\sqrt{-g}}\frac{\delta S}{\delta g_{\alpha\beta}}, \qquad\qquad\pi^{\alpha}=-\frac{16\pi G_{D+2}}{\sqrt{-g}}\frac{\delta S}{\delta\CA_\alpha}.
\eea
Equivalently, the variation of the on-shell action is:
\bea
\delta S &=& -\frac{1}{16\pi G_{D+2}}\int_{\partial\CM}d^dx 
\sqrt{-g}\left[ \pi^{\alpha\beta}\delta g_{\alpha\beta} + \pi^\alpha\delta\CA_\alpha\right]\\
\label{pi2}
&=&-\frac{1}{16\pi G_{D+2}}\int_{\partial\CM} d^dx \sqrt{-g}\left[ (2{\pi^\alpha}_\beta + \pi^\alpha\CA_\beta)
e^\beta_{B} \delta e^{B}_\alpha + \pi^A \delta\CA_A\right].
\eea
The boundary stress tensor ${T^\alpha}_B$, however, is defined by functional differentiation of the on-shell action with respect to the vielbeins $e_\alpha^B$, while holding the vector field with $\textit{frame indices}$ ($\CA_A $) fixed. Note that $\CA_0=\alpha+\psi$ is the only non-zero component of $\CA_A$. Therefore, the variation of the on-shell action can also be written as:
\bea
\label{st2}
\delta S = -\frac{1}{16\pi G_{D+2}}\int d^dx \sqrt{\gamma}N\left[ {T^\alpha}_B
\delta e^{B}_\alpha + \pi_\psi \delta\psi\right],
\eea
where
\bea
\label{stresstensor}
{T^A}_B&=&-\frac{16\pi G_{D+2}}{\sqrt{\gamma}N}e_\alpha^A\frac{\delta S}{\delta e_\alpha^B}=-\frac{1}{\sqrt{\gamma}N}e_\alpha^A\frac{\delta}{\delta e_\alpha^B}
\int dt\,d^D\bx\,\sqrt{\gamma}\,N\CL,\\
\label{masvector}
\pi_\psi&=&-\frac{16\pi G_{D+2}}{\sqrt{\gamma}N}\frac{\delta S}{\delta \psi}=-\frac{1}{\sqrt{\gamma}N}\frac{\delta}{\delta \psi}\int dt\,d^D\bx\,\sqrt{\gamma}\,N\CL.
\eea
By comparing equations (\ref{pi2}) and (\ref{st2}) we get the following relations:
\be
T_{\alpha B}=(2\pi_{\alpha\beta}+\pi_\alpha\CA_\beta)e^\beta_B, \qquad \qquad \pi_\psi=\pi^0.
\ee
Rearranging these expressions we have 
\bea
\label{pitrel}
\pi_{AB}=\frac{1}{2}(T_{AB}-\pi_A\CA_B), \qquad\qquad \pi_I\CA_0=T_{I0}-T_{0I}.
\eea

Finally, by using the expressions for the vielbeins derived in Appendix~\ref{apadm}, 
we can write the stress tensor as:
\bea
{T^0}_0&=&-\frac{16\pi G_{D+2}}{\sqrt{\gamma}}\frac{\delta S}{\delta N},\\
{T^0}_I&=&-\frac{16\pi G_{D+2}}{\sqrt{\gamma}}\frac{\delta S}{\delta N^I},\\
{T^I}_J&=&-16\pi G_{D+2}\left(\frac{N^I}{\sqrt{\gamma}N}\frac{\delta S}{\delta N^J}+\frac{1}{\sqrt{\gamma}N}\hat{e}_i^I\frac{\delta S}{\delta \hat{e}_i^J}\right)\cr
&=&-16\pi G_{D+2}\left(\frac{N^I}{\sqrt{\gamma}N}\frac{\delta S}{\delta N^J}+\frac{2}{\sqrt{\gamma}N}\hat{e}_i^I\hat{e}_{Jj}\frac{\delta S}{\delta \gamma_{ij}}\right).
\eea
We will use these expressions to determine the stress tensor and vector momentum from the on-shell action

\subsection{Boundary source fields and asymptotic scaling}
\label{appbsfaas}

The boundary conditions are specified by fixing the sources for the various field theory operators on the boundary. Our boundary conditions involve the following finite fixed sources as $r\rightarrow\infty$ (denoting each source with a bar):
\bea
\bar{e}_\alpha^0=\frac{e_\alpha^0}{r^z}, \qquad \bar{e}_\alpha^I=\frac{e_\alpha^I}{r}, \qquad \bar{\psi}=\frac{\psi}{r^{-\D_-}}. 
\eea
In order to have a foliation on the boundary, it is necessary to set  $\bar{e}_i^0$ (the source for the energy flux $\mathcal{E}^i$) equal to zero \cite{ross}. For all of this paper, we have set  $\bar{e}_i^0=0$.

Note that ${T^\alpha}_A$ is the vacuum expectation value of the operator sourced by $\bar{e}_\alpha^A$. In other words, $\bar{e}_\alpha^0$ is the source for the energy density $\mathcal{E}$ and the energy flux $\mathcal{E}^i$, whereas $\bar{e}_\alpha^I$ is the source for the momentum density $\mathcal{P}_i$ and the stress tensor  ${{\Pi}^i}_j$. $\bar{\psi}$ is the source for $\CO_\psi$, the operator dual to the massive vector $\psi$.
The operator $\CO_\psi$ is relevant for $z<D$ and irrelevant for $z>D$. Therefore, for $z>D$, we must take $\bar{\psi}=0$ in order to preserve the asymptotic boundary conditions above. In the case $z=D$, the operator is 
classically marginal, and there is some evidence suggesting that it becomes 
marginally relevant in the case of $D=2$ \cite{xi}.

Note also that the scaling dimensions discussed here are the classical scaling 
dimensions, consistent with the fact that we perform our analysis near the 
ultraviolet fixed point with fixed $z$.  In the bulk, this corresponds to the 
asymptotic analysis in the vicinity of the spacetime boundary at conformal 
infinity.  Hence, in our analysis we systematically ignore most of the 
possible nontrivial infrared dynamics, such as the flow -- generically 
expected of Lifshitz-type theories -- towards lower values of $z$ under the 
influence of relevant operators.

The above scaling behavior allows us to determine the scaling behavior of other quantities near the boundary. Any boundary quantity can be written in terms of the source fields $\bar{e}_\alpha^A$ and $\bar{\psi}$ and then the scaling behavior can be read off from the resulting exponents of $r$.
Consider a general object $\CO$. When written in terms of the boundary source fields, we say that the term  in $\CO$ scaling as $r^{-\D}$ is of ``order $\D$'' and denote it by $\CO^{(\D)}$. For example, $e_\alpha^0$ has order $-z$, $e_\alpha^I$ has order $-1$ and $\psi$ has order $\D_-$. This means that $N$ has order $-z$, $N_i$ has order $-2$, $\gamma_{ij}$ has order $-2$ and $\gamma^{ij}$ has order $2$.

From equation (\ref{riccidec}), $R$ has components of order $2$ and $2z$ given by:
\bea
R^{(2)}&=&\hat{R}-\frac{2\nabla^i\nabla_iN}{N},\\
R^{(2z)}&=&\hat{K}_{ij}\hat{K}^{ij}-\hat{K}^2 +\frac{\partial_tZ}{N\sqrt{\gamma}}
+\frac{\nabla^iY_i}{N}.
\eea
From equation (\ref{fsquareddec}), $F_{AB}F^{AB}$ has components of order $2$, $2+\D_-$ and $2+2\D_-$ given by:
\bea
(F_{AB}F^{AB})^{(2)}&=&-\frac{2\alpha^2\nabla_i N\nabla^i N}{N^2},\\
(F_{AB}F^{AB})^{(2+\D_-)}&=& -\frac{4\alpha \nabla_i (N\psi)\nabla^i N}{N^2},\\
(F_{AB}F^{AB})^{(2+2\D_-)}&=& -\frac{2\nabla_i (N\psi)\nabla^i (N\psi) }{N^2}.
\eea
Also,  $\CA_A\CA^A=-(\alpha+\psi)^2$ has components of dimension $0,\D_-, 2\D_-$:
\bea
(\CA_A\CA^A)^{(0)}&=&-\alpha^2,\\
(\CA_A\CA^A)^{(\D_-)}&=& -2\alpha\psi,\\
(\CA_A\CA^A)^{(2\D_-)}&=& -\psi^2.
\eea

Note also that equations (\ref{stresstensor}) and (\ref{masvector}) imply that terms $\CL^{(\D)}$ in the on-shell action determine ${{T^0}_0}^{(\D)}$, ${{T^0}_I}^{(\D+1-z)}$, ${{T^I}_0}^{(\D+z-1)}$, ${{T^I}_J}^{(\D)}$ and ${\pi_\psi}^{(\D-\D_-)}$.

\section{Holographic Renormalization Equations}
\label{apphre}

The on-shell action is a function of the boundary fields and is written as
\be
\label{onshacti}
S=\frac{1}{16\pi G_{D+2}}\int dt\,d^D\bx\,\sqrt{\gamma}\,N\CL.
\ee

A convenient way of computing the divergent part of $\CL$  is to organize the terms with respect to how they scale with $r$. More precisely, we define the dilatation operator by:
\be
\deltaD = \int dt\,d^D\bx\, \left(
z e^0_\mu \frac{\delta}{\delta e^0_\mu}
+ e^I_\mu \frac{\delta}{\delta e^I_\mu}
-\D_- \psi \frac{\delta}{\delta \psi}
\right) .
\ee
This operator asymptotically represents $\displaystyle r\frac{\p}{\p r}$.

$\CL$ can then be decomposed into a sum of terms as follows:
\be
\label{llexpi}
\CL = \sum_{\Delta\geq0} \CL^{(\Delta)} + \tilde\CL^{(z+D)}\log r.
\ee
Note that we include a logarithmic term at order $z+D$ due to the possibility of a Weyl scaling anomaly. The individual terms of the expansion (\ref{llexpi}) satisfy
\be
\deltaD\CL^{(\Delta)} = -\Delta\, \CL^{(\Delta)} 
\qquad {\rm for}\ \ \Delta \ne z+D,
\ee
\be
\deltaD\CL^{(z+D)} = -(z+D)\CL^{(z+D)} + \tilde\CL^{(z+D)},
\ee
\be
\deltaD\tilde\CL^{(z+D)} =- (z+D)\tilde\CL^{(z+D)}.
\ee
Applying $\deltaD$ to the on-shell action (\ref{onshacti}) and using equations (\ref{stresstensor}) and (\ref{masvector}) then yields:
\be
(z+D+\deltaD)\CL = -z{T^0}_0-{T^I}_I +\D_-\psi\pi_\psi.
\ee
Expanding this at each order then results in:
\be
\label{divterms}
(z+D-\Delta)\CL^{(\Delta)} = -z{{T^0}_0}^{(\Delta)} - {{T^I}_I}^{(\Delta)} + \D_-\psi\pi_\psi^{(\Delta-\D_-)}
\ee
except for $\Delta=z+D$, when this becomes
\be
 \tilde\CL^{(z+D)} =-z{{T^0}_0}^{(z+D)} - {{T^I}_I}^{(z+D)} + 
\D_-\psi\pi_\psi^{(z+D-\D_-)}.
\ee
This allows us to solve for the anomaly. The above equations imply that the anomaly term can also be found by:
\be
\label{anomres}
 \tilde\CL^{(\Delta)} =\lim_{\Delta\rightarrow z+D}\left((z+D-\Delta)\CL^{(\Delta)}\right).
\ee
Note that the value of $\CL^{(z+D)}$ cannot be found by this asymptotic analysis.

We now move on to finding an explicit expression for these divergent terms in the onshell action $\CL^{(\Delta)}$. The variation of the bulk action (\ref{baction}) with respect to $\CN$ produces the Hamiltonian constraint equation,
\bea
K^2-K_{AB}K^{AB}-\frac{1}{2}\pi_A\pi^A-\frac{1}{2m^2}(\nabla^A\pi_A)^2=R-2\Lambda-\frac{1}{4}F_{AB}F^{AB}-\frac{1}{2}m^2\CA_A\CA^A.
\eea
Expanding this equation in its dilatation eigenvalues (utilizing equations (\ref{pitrel}), (\ref{k0calc}), (\ref{pi0calc}), (\ref{piacalc})) and then substituting it into equation (\ref{divterms}) yields an expression for $\CL^{(\D)}$ (see \cite{ross} for more details). 
Explicitly, the terms in the on-shell action are given for $\D\neq 0$, $\D_-$ 
and $2\D_-$ by:
\bea
(z+D-\D)\CL^{(\D)}=\FQ^{(\D)}+\FS^{(\D)},
\eea
where the quadratic term $\FQ^{(\Delta)}$ is given by
\bea
\FQ^{(\D)}&=&\sum_{0<s<\D/2;s\neq\D_-}\left[2K_{AB}^{(s)}\pi^{AB(\D-s)}
+\pi_A^{(s)}\pi^{A(\D-s)}+\frac{1}{m^2}(\nabla_A\pi^A)^{(s)}
(\nabla_A\pi^A)^{(\D-s)}\right]\cr
&&\qquad\qquad{}+\left[\vphantom{\frac{1}{2}}
K_{AB}^{(\D_-)}T^{AB(\D-\D_-)}+K_{00}^{(\D_-)}
\pi^{0(\D-2\D_-)}\psi+\pi_I^{(\D_-)}\pi^{I(\D-\D_-)}\right]\cr
&&\qquad\qquad\qquad{}+\left[K_{AB}^{(\D/2)}\pi^{AB(\D/2)}+\frac{1}{2}
\pi_A^{(\D/2)}\pi^{A(\D/2)}+\frac{1}{2m^2}(\nabla_A\pi^A)^{(\D/2)2}\right]
\eea
and the source $\FS$ is
\be
\FS=R-2\Lambda-\frac{1}{4}F_{AB}F^{AB}-\frac{1}{2}m^2\CA_A\CA^A.
\ee
We also have the following exceptions to the above formula:
\bea
(z+D)\CL^{(0)}&=&2 \FS^{(0)},\\
(z+D-\D_-)\CL^{(\D_-)}&=&(\D_--z)\psi\pi_\psi^{(0)}+ \FS^{(\D_-)},\\
(z+D-2\D_-)\CL^{(2\D_-)}&=&(\D_--z)\psi\pi_\psi^{(\D_-)}+K_{AB}^{(\D_-)}\pi^{AB(\D_-)}\cr
&&\qquad\qquad\qquad{}+\frac{1}{2}\pi_A^{(\D_-)}\pi^{A(\D_-)}+\FS^{(2\D_-)}.
\eea

$\FS$ needs to be calculated at each order. The calculation in Appendix~\ref{appbsfaas} shows that $R$ has components of order $2$ and $2z$, $F_{AB}F^{AB}$ has components of order $2, 2+\D_-, 2+2\D_-$ and $\CA_A\CA^A$ has components of order $0,\D_-, 2\D_-$, resulting in:
\bea
\FS^{(0)}&=&-2\Lambda+\frac{1}{2}m^2\alpha^2=(z+D)(z+D-1),\\
\FS^{(\D_-)}&=&m^2\alpha\psi=Dz\alpha\psi,\\
\FS^{(2\D_-)}&=&\frac{1}{2}m^2\psi^2=\frac{Dz}{2}\psi^2,\\
\FS^{(2)}&=&R^{(2)}-\frac{1}{4}(F_{AB}F^{AB})^{(2)}=\hat{R}
-2\frac{\nabla^2N}{N}+\frac{\alpha^2}{2}\left(\frac{\nabla N}{N}\right)^2,\\
\FS^{(2+\D_-)}&=&-\frac{1}{4}(F_{AB}F^{AB})^{(2+\D_-)}=\frac{\alpha\nabla_iN\nabla^i(N\psi)}{N^2},\\
\FS^{(2+2\D_-)}&=&-\frac{1}{4}(F_{AB}F^{AB})^{(2+2\D_-)}=\frac{\nabla^i(N\psi)\nabla_i(N\psi)}{2N^2},\\
\FS^{(2z)}&=&R^{(2z)}=\hat{K}_{ij}\hat{K}^{ij}-\hat{K}^2 + \textrm{total derivatives}.
\eea
We now proceed to use these formulae to calculate the divergent terms in the on-shell action at each order. Once these divergent terms have been calculated, counterterms must be added to the action in order to subtract these divergences.  
With a boundary cutoff at $\displaystyle r=\frac{1}{\epsilon}$), the 
counterterms are
\bea
S_{ct}=-\frac{1}{16\pi G_{D+2}}\int dt\,d^D\bx\,\sqrt{\gamma}\,N\left(
\sum_{0\leq\Delta<z+D} \CL^{(\Delta)} - \tilde\CL^{(z+D)}\log \epsilon\right).
\eea

\subsection{Non-derivative counterterms}
At order $0$, we have:
\bea
\CL^{(0)}&=&\frac{2\FS^{(0)}}{z+D}=2(z+D-1).
\eea
This yields ${{T^A}_B}^{(0)}=-2(z+D-1){\delta^A}_B$.

To evaluate the order $\D_-$ and $2\D_-$ counterterms we need some additional information. From the asymptotic expansions given in \cite{ross}, it is clear that:
\bea
\label{k0calc}
{{K^0}_0}^{(0)}=z, \qquad {{K^I}_J}^{(0)}={\delta^I}_J.
\eea
Also, the zero-component of the vector momentum is given by:
\bea
\pi_0=rF_{r0}=r\partial_r{\CA}_0+\CA_0{K^0}_0-r\partial_0\CA_r.
\eea
This gives:
\bea
\label{pi0calc}
\pi_0^{(0)}&=&\alpha{{K^0}_0}^{(0)}=\alpha z\\
\label{piacalc}
\pi_0^{(\D_-)}&=&r\partial_r{\psi}+\alpha{{K^0}_0}^{(\D_-)}+\psi{{K^0}_0}^{(0)}=\alpha{{K^0}_0}^{(\D_-)}+(z-\D_-)\psi
\eea
Note that $\pi_\psi\equiv\pi^0$.

Then:
\bea
(z+D-\D_-)\CL^{(\D_-)}&=&-(z-\D_-)\psi\pi_\psi^{(0)}+ \FS^{(\D_-)}=(z-\D_-)\psi\alpha z+Dz\alpha\psi\\
\CL^{(\D_-)}&=&z\alpha\psi
\eea
which yields ${{T^A}_B}^{(\D_-)}=-z\alpha\psi{\delta^A}_B$. 

Note that ${\pi^A}_B=\frac{1}{2}({T^A}_B-\pi^A\CA_B)={K^A}_B-K{\delta^A}_B$ and this means that:
\bea
{{\pi^0}_0}^{(\D_-)}&=&\frac{1}{2}({{T^0}_0}^{(\D_-)}-\alpha\pi_\psi^{(\D_-)}-\psi\pi_\psi^{(0)})=-\frac{1}{2}\alpha\pi_\psi^{(\D_-)}\\
{{\pi^I}_J}^{(\D_-)}&=&\frac{1}{2}{{T^I}_J}^{(\D_-)}=-\frac{z\alpha\psi}{2}{\delta^I}_J\\
K^{(\D_-)}&=&-\frac{{{\pi^A}_A}^{(\D_-)}}{D}=\frac{\alpha\pi_\psi^{(\D_-)}}{2D}+\frac{z\alpha\psi}{2}\\
{{K^0}_0}^{(\D_-)}&=&{{\pi^0}_0}^{(\D_-)}+K^{(\D_-)}=-\frac{\alpha\pi_\psi^{(\D_-)}(D-1)}{2D}+\frac{z\alpha\psi}{2}\\
{{K^I}_J}^{(\D_-)}&=&{{\pi^I}_J}^{(\D_-)}+K^{(\D_-)} {\delta^I}_J=\frac{\alpha\pi_\psi^{(\D_-)}}{2D}{\delta^I}_J
\eea
Substituting this into the expression $\pi_0^{(\D_-)}=\alpha{{K^0}_0}^{(\D_-)}+(z-\D_-)\psi$ derived above gives:
\bea
\pi_0^{(\D_-)}&=&\alpha(-\frac{\alpha\pi_\psi^{(\D_-)}(D-1)}{2D}+\frac{z\alpha\psi}{2})+(z-\D_-)\psi\\
\pi_0^{(\D_-)}&=&\frac{2D(2z-1-\D_-)}{2D-\alpha^2(D-1)}\psi=\frac{Dz(2z-1-\D_-)}{z+D-1}\psi\\
\eea
Therefore, using this result for $\pi_0^{(\D_-)}$:
\bea
K^{(\D_-)}&=&-\frac{\alpha z (2z-1-\D_-)}{2(z+D-1)}\psi+\frac{z\alpha\psi}{2}=-\frac{\alpha z(z-D-\D_-)}{2(z+D-1)}\psi\\
{{K^0}_0}^{(\D_-)}&=&\frac{\alpha z(D-1)(2z-1-\D_-)}{2(z+D-1)}\psi+\frac{z\alpha\psi}{2}=\frac{\alpha z((2D-1)z-(D-1)\D_-)}{2(z+D-1)}\psi\\
{{K^I}_J}^{(\D_-)}&=&-\frac{\alpha z(2z-1-\D_-)}{2(z+D-1)}\psi{\delta^I}_J
\eea
Then:
\bea
(z+D-2\D_-)\CL^{(2\D_-)}&=&-(z-\D_-)\psi\pi_\psi^{(\D_-)}+K_{AB}^{(\D_-)}\pi^{AB(\D_-)}+\frac{1}{2}\pi_A^{(\D_-)}\pi^{A(\D_-)}+ \FS^{(2\D_-)}\cr
&=&-(z-\D_-)\psi\pi_\psi^{(\D_-)}+(-\frac{\alpha\pi_\psi^{(\D_-)}(D-1)}{2D}+\frac{z\alpha\psi}{2})(-\frac{1}{2}\alpha\pi_\psi^{(\D_-)})\cr
&&+(\frac{\alpha\pi_\psi^{(\D_-)}}{2})(-\frac{z\alpha\psi}{2})+\frac{1}{2}(\pi_\psi^{(\D_-)})^2- \frac{Dz}{2}\psi^2\cr
&=&\frac{Dz\psi^2(4z^2-4z-4z\D_-+1+2\D_-+\D_-^2+z+D-1)}{2(z+D-1)}\cr
&=&\frac{Dz\psi^2(z + D - 2 \D_-) (2 z - 1 - \D_-)}{2(z+D-1)}\\
\CL^{(2\D_-)}&=&\frac{Dz\psi^2(2 z - 1 - \D_-)}{2(z+D-1)}
\eea
where $\D_- = \frac{1}{2}(z + D - \beta_z)$ and $\beta_z = \sqrt{(z + D)^2 + 8 (z - 1)(z - D)}$ has been used.

This result yields $\displaystyle {{T^A}_B}^{(2\D_-)}=-\frac{Dz\psi^2(2 z - 1 - \D_-)}{2(z+D-1)}{\delta^A}_B$. Next we can calculate:
\bea
(z+D-3\D_-)\CL^{(3\D_-)}&=&K_{AB}^{(\D_-)}T^{AB(2\D_-)}+K_{00}^{(\D_-)}\pi^{0(0)}
\psi\cr
&=&\frac{D\alpha z^2(2 z - 1 - \D_-)(z-D-\D_-)}{4(z+D-1)^2}\psi^3\cr
&&+\frac{D\alpha z^2((2D-1)z-(D-1)\D_-)(2z-1-\D_-)}{2(z+D-1)^2}\psi^3\cr
&=&\frac{D\alpha z^2(2 z - 1 - \D_-)(-D+(4D-1)z-(2D-1)\D_-)}{4(z+D-1)^2}\psi^3
\nonumber
\eea
which yields $\displaystyle \pi_\psi^{(2\D_-)}=-\frac{3D\alpha z^2(2 z - 1 - \D_-)(-D+(4D-1)z-(2D-1)\D_-)}{4(z+D-3\D_-)(z+D-1)^2}\psi^2$. 

This allows us to calculate $K_{AB}^{(2\D_-)}$:
\bea
{{\pi^0}_0}^{(2\D_-)}&=&\frac{1}{2}({{T^0}_0}^{(2\D_-)}-\alpha\pi_\psi^{(2\D_-)}-\psi\pi_\psi^{(\D_-)})=-\frac{1}{2}\alpha\pi_\psi^{(2\D_-)}-\frac{1}{4}\psi\pi_\psi^{(\D_-)}\\
{{\pi^I}_J}^{(2\D_-)}&=&\frac{1}{2}{{T^I}_J}^{(2\D_-)}=\frac{1}{4}\psi\pi_\psi^{(\D_-)}{\delta^I}_J\\
K^{(2\D_-)}&=&-\frac{{{\pi^A}_A}^{(\D_-)}}{D}=\frac{1}{2D}\alpha\pi_\psi^{(2\D_-)}-\frac{(D-1)}{4D}\psi\pi_\psi^{(\D_-)}\\
{{K^0}_0}^{(2\D_-)}&=&{{\pi^0}_0}^{(2\D_-)}+K^{(2\D_-)}=-\frac{\alpha\pi_\psi^{(2\D_-)}(D-1)}{2D}-\frac{(2D-1)}{4D}\psi\pi_\psi^{(\D_-)}\\
{{K^I}_J}^{(2\D_-)}&=&{{\pi^I}_J}^{(2\D_-)}+K^{(2\D_-)} {\delta^I}_J=\left(\frac{1}{2D}\alpha\pi_\psi^{(2\D_-)}+\frac{1}{4D}\psi\pi_\psi^{(\D_-)}\right){\delta^I}_J
\eea

Higher order non-derivative terms can be calculated in a similar manner.

\subsection{Two-derivative counterterms with $\psi=0$}

Up to total derivatives, the divergent term in the on-shell action of order 2 
is:
\bea
(z+D-2)\CL^{(2)}&=&\FS^{(2)}=\hat{R}-\frac{2\nabla^i\nabla_iN}{N}
+\frac{\alpha^2\nabla^iN\nabla_iN}{2N^2}=\hat{R}+\frac{\alpha^2\nabla^iN
\nabla_iN}{2N^2}
\eea
This gives the following contribution to the stress tensor (see 
Appendix~\ref{ape}):
\bea
(z+D-2)T_{00}^{(2)}&=&\hat{R}-\frac{\alpha^2\nabla^i\nabla_iN}{N}
+\frac{\alpha^2\nabla^iN\nabla_iN}{2N^2},\cr
(z+D-2)T_{0I}^{(3-z)}&=&0,\cr
(z+D-2)T_{IJ}^{(2)}&=&2\hat{R}_{IJ}-\frac{2\nabla_I\nabla_JN}{N}
+\frac{\alpha^2\nabla_IN\nabla_JN}{N^2},\cr
&&\qquad\qquad\qquad{}+\delta_{IJ}\left(-\hat{R}+\frac{2\nabla^i\nabla_iN}{N}
-\frac{\alpha^2\nabla^iN\nabla_iN}{2N^2}\right),\cr
(z+D-2){{T^I}_I}^{(2)}&=&-(D-2)\hat{R}+\frac{2(D-1)\nabla^i\nabla_iN}{N}
-\frac{\alpha^2(D-2)\nabla^iN\nabla_iN}{2N^2}.\nonumber
\eea
At order $2z$ there is a contribution from the quadratic term $\displaystyle\frac{1}{2m^2}(\nabla_A\pi^A)^{(z)2}$. Note that:
\bea
(\nabla_A\pi^A)^{(z)}&=&(\partial_A\pi^A-{\omega_{A}}^{AB}\pi_B)^{(z)}
=(\partial_A\pi^A-2{\Omega_{A}}^{BA}\pi_B)^{(z)}\cr
&=&(\partial_0(\pi^{0(0)})-2{\Omega_{I}}^{0I}\pi_0^{(0)})
=(\partial_0(-z\alpha)-2{\Omega_{I}}^{0I}z\alpha)\cr
&&\qquad\qquad=-\alpha z \hat{K},
\eea
where expressions from Appendix~\ref{apadm} have been used.
Therefore, up to total derivatives:
\bea
(z+D-2z)\CL^{(2z)}&=&\FS^{(2z)}+\frac{1}{2m^2}(\nabla_A\pi^A)^{(z)2}\cr
&=&\hat{K}_{ij}\hat{K}^{ij}-\hat{K}^2+\frac{1}{2m^2}(-\alpha z \hat{K})^2\cr
&=&\hat{K}_{ij}\hat{K}^{ij}-\frac{(1+D-z)}{D}\hat{K}^2
\eea

\subsection{Two-derivative counterterms involving $\psi$}

We can also calculate various divergent terms involving $\psi$, for example:
\bea
&&(z+D-2-\D_-)\CL^{(2+\D_-)}=K_{AB}^{(\D_-)}T^{AB(2)}+\frac{\alpha\nabla_iN\nabla^i(N\psi)}{N^2}\cr
&&\quad{}=-\frac{\alpha z \psi}{2(z+D-1)}\left[ ((2D-1)z-(D-1)\D_-)T^{00(2)}+(2z-1-\D_-){{T^I}_I}^{(2)}\right]\cr
&&\qquad\qquad\qquad\qquad-\frac{\alpha\nabla^i\nabla_iN\psi}{N}+\frac{\alpha\nabla_iN\nabla^iN\psi}{N^2}\cr
&&\quad=-\frac{\alpha z \psi}{2(z+D-1)(z+D-2)}\left[\left((2D-1)z-(D-1)\D_-\right)\left(\hat{R}-\frac{\alpha^2\nabla^i\nabla_iN}{N}+\frac{\alpha^2\nabla^iN\nabla_iN}{2N^2}\right)\right.\cr
&&\qquad\qquad\left.{}+(2z-1-\D_-)\left(-(D-2)\hat{R}+\frac{2(D-1)\nabla^i\nabla_iN}{N}-\frac{\alpha^2(D-2)\nabla^iN\nabla_iN}{2N^2}\right)\right]\cr
&&\qquad\qquad\qquad\qquad-\frac{\alpha\nabla^i\nabla_iN\psi}{N}+\frac{\alpha\nabla_iN\nabla^iN\psi}{N^2}.
\eea
Or, by defining some constants:
\bea
\CL^{(2+\D_-)}&=&-\psi\left(c_1\hat{R}+c_2\frac{\nabla^i\nabla_iN}{N}+c_3\frac{\nabla_iN\nabla^iN}{N^2}\right)
\eea
where:
\bea
c_1&=&\frac{\alpha z(-2+D-\D_-+3z)}{2(D-2 + z)(D-1+z)(z+D-2-\D_-)}\cr
c_2&=&\frac{\alpha (4+2D^2-(4+(\alpha^2-2) \D_-) z + ( \alpha^2-2) z^2+ 
   D ((2 + (\alpha^2-2) \D_-) z - 2 (\alpha^2-2) z^2-6))}{2 (D-2 +
    z) (D-1 + z)(z+D-2-\D_-)}\cr
c_3&=&\frac{\alpha (-8 - 4 D^2 -(-12+\alpha^2(2+\D_-))z+(3\alpha^2-4) z^2 + 
D(12+(\alpha^2-8)z))}{4(D-2+z)(-1+D+z)(-2+D-\D_-+z)}\nonumber
\eea
(Note that for $z=D=2$ we have $c_1=\frac{1}{2}$, $c_2=\frac{1}{2}$ and 
$c_3=-\frac{1}{4}$.)

This results in:
\bea
\pi_\psi^{(2)}&=&c_1\hat{R}+c_2\frac{\nabla^i\nabla_iN}{N}+c_3\frac{\nabla_iN\nabla^iN}{N^2}
\eea
and also:
\bea
T_{00}^{(2+\D_-)}&=&-c_1\psi\hat{R}-c_2\nabla^i\nabla_i\psi-c_3\frac{\nabla_iN\nabla^iN\psi}{N^2}+2c_3\frac{\nabla_i\nabla^iN\psi}{N}+2c_3\frac{\nabla^iN\nabla_i\psi}{N}\\
T_{0I}^{(3-z+\D_-)}&=&0\\
T_{IJ}^{(2+\D_-)}&=&\delta_{IJ}\left(c_1\psi\hat{R}-c_2\frac{\nabla_iN\nabla^i\psi}{N}+c_3\frac{\nabla_iN\nabla^iN\psi}{N^2}\right)\cr
&&-2c_1\psi\hat{R}_{IJ}+c_2\frac{\nabla_IN\nabla_J\psi}{N}+c_2\frac{\nabla_JN\nabla_I\psi}{N}-2c_3\frac{\nabla_IN\nabla_JN\psi}{N^2}\cr
&&-2\delta_{IJ}c_1\frac{\nabla^i\nabla_i(N\psi)}{N}+2c_1\frac{\nabla_I\nabla_J(N\psi)}{N}\\
{{T^I}_I}^{(2+\D_-)}&=&(D-2)\left(c_1\psi\hat{R}-c_2\frac{\nabla_iN\nabla^i\psi}{N}+c_3\frac{\nabla_iN\nabla^iN\psi}{N^2}\right)-2c_1\frac{\nabla^i\nabla_i(N\psi)}{N}
\eea
There are many more two-derivative terms involving $\psi$. For example:
\bea
(z+D-2-2\D_-)\CL^{(2+2\D_-)}&=&2K_{AB}^{(2\D_-)}\pi^{AB(2)}+\pi_A^{(2\D_-)}\pi^{A(2)}\cr
&+&K_{AB}^{(\D_-)}T^{AB(2+\D_-)}+K_{00}^{(\D_-)}\pi^{0(2)}\psi
+\FS^{(2+2\D_-)}
\eea
This has been calculated explicitly in the case $D=z=2$:
\bea
\CL^{(2+2\D_-)}&=&\psi^2(\frac{\nabla^i\nabla_iN}{8N}-\frac{\nabla^iN\nabla_iN}{2N^2})+\frac{3}{4}\psi\nabla^i\nabla_i\psi
\eea

\subsection{Four-derivative counterterms with $\psi=0$}

At fourth order we have:
\bea
&&(z+D-4)\CL^{(4)}=K_{AB}^{(2)}\pi^{AB(2)}+\frac{1}{2}\pi_A^{(2)}\pi^{A(2)}\cr
&&\qquad=\frac{1}{a_0}[a_1(\frac{\nabla_iN\nabla^iN}{N^2})^2  +a_2\frac{\nabla_iN\nabla^iN}{N^2}\hat{R} + a_3 \frac{\nabla^i\nabla_iN}{N} \frac{\nabla_jN\nabla^jN}{N^2}\cr
&&\qquad\qquad\quad+ a_4\frac{\nabla^iN\nabla^jN}{N^2} \hat{R}_{ij} +  a_5 \frac{\nabla^i\nabla_iN}{N} \hat{R} + a_6(\frac{\nabla^i\nabla_iN}{N})^2 + a_7 \hat{R}_{ij}^2 + a_8 \hat{R}^2]
\eea
where:
\bea
a_0 &=& -2 D z^2 (-2 + D + z)^2 (-1 + D + z) (-4 + \beta + D + z)^2\\
a_1&=&  32 (z-1)^3 + D^4 (-11 + z (6 + z)) \cr
&&+  D^3 (52 - 3 \beta_z - z (77 + 2 \beta_z - (34 + \beta_z) z + z^2)) \cr
&&+  D^2 (16 (-8 + \beta_z) +  z (2 (116 + \beta_z) +  z (-145 - 8 \beta_z + 2 (9 + \beta_z) z + 3 z^2))) \cr
&&+  D (z-1) (16 (-8 + \beta_z) +  z (184 + z (-68 - 5 \beta_z + z (-13 + \beta_z + 5 z)))) \\
a_2&=&2 z (z-1) (D^4 + D^3 (\beta_z - z) + 16 (z-1) z +  D^2 (8 - 4 \beta_z + z (-16 + 2 \beta_z + 3 z)) \cr
&&+  D z (-4 (-8 + \beta_z) + z (-24 + \beta_z + 5 z)))      \\
a_3&=&     -2 z (D^4 (z-4) + 32 (z-1)^2 + D^3 (-2 (7 + 2 \beta_z) + (21 + \beta_z - z) z) \cr
&&+  D^2 (32 + 18 \beta_z + z (-60 - 11 \beta_z + z (10 + 2 \beta_z + 3 z)))\cr
&&+  D (z-1) (8 (8 + \beta_z) +  z (-40 - 6 \beta_z + z (-18 + \beta_z + 5 z))))  \\
a_4&=&  -4 z  D (z-2) (D-1 + z) (8 + D^2 - 8 z - 2 D z + 5 z^2 + \beta_z (-4 + D + z))\\
a_5&=&       -  4  z^2 (D^3 + D^2 (\beta_z - 2 z) + 8 (z-1) z +  D (8 + \beta_z (z-4 ) - 3 z^2)) \\
a_6&=&   - 4 z (D-1) (D^3 + D^2 (\beta_z - 2 z) + 8 (z-1) z + D (8 + \beta_z (z-4 ) - 3 z^2)) \\
a_7&=&   -4 z^2  D (D-1 + z) (8 + D^2 - 8 z - 2 D z + 5 z^2 +  \beta_z (D-4 + z)) \\
a_8&=&  z^2 (D^4 + D^3 (\beta_z - z) + 8 (z-1) z^2 +  D^2 (8 - 4 \beta_z + z (-8 + 2 \beta_z + 3 z)) \cr
&&+  D z (8 - 4 \beta_z + z (-8 + \beta_z + 5 z)))
\eea
In the above expression we have used the following identities for terms in the action (up to total derivatives):
\bea
\frac{\nabla_iN\nabla_jN\nabla^i\nabla^jN}{N^3}&\sim&\left(\frac{\nabla_iN
\nabla^iN}{N^2}\right)^2-\frac{\nabla_iN\nabla^iN\nabla^j\nabla_jN}{2N^3}\cr
\frac{\nabla_i\nabla_jN\nabla^i\nabla^jN}{N^2}&\sim& \left(\frac{\nabla_iN\nabla^iN}{N^2}\right)^2 - \frac{3\nabla_iN\nabla^iN\nabla^j\nabla_jN}{2N^3} + \left(\frac{\nabla^i\nabla_iN}{N}\right)^2 - \frac{\nabla^iN\nabla^jN R_{ij}}{N^2}\cr
\frac{\nabla^i\nabla^jN R_{ij}}{N}&\sim&\frac{\nabla^i\nabla_iN R}{2N}\nonumber
\eea

For $D=2$ we have further simplifications because $\displaystyle R_{ij}=\frac{R}{2}\delta_{ij}$ and so:
\bea
(z-2)\CL^{(4)}&=&\frac{(z-2)}{b_0}\left[b_1\left(\frac{\nabla_iN\nabla^iN}{N^2}\right)^2  +b_2\frac{\nabla_iN\nabla^iN}{N^2}\hat{R} + b_3 \frac{\nabla^i\nabla_iN}{N} \frac{\nabla_jN\nabla^jN}{N^2}\right. \\
&&\qquad\left.{}+b_4 \frac{\nabla^i\nabla_iN}{N} \hat{R} + b_5\left(\frac{\nabla^i\nabla_iN}{N}\right)^2 + b_6 \hat{R}^2\right],
\eea
where:
\bea
b_0&=&-2z^4(z+1)(z-2+\beta_z)^2\\
b_1&=&12 + 36 z - 11 z^2 - 2 z^3 + 5 z^4 + \beta_z (-2 - 7 z + z^3)\\
b_2&=&4 z^2 (z-6 + \beta_z)\\
b_3&=&-2 z (36 - 4 z - 7 z^2 + 5 z^3 + \beta_z (z^2- z -6))\\
b_4&=&-4 z^2 (z-6 + \beta_z)\\
b_5&=&-4 z (z-6 + \beta_z)\\
b_6&=&-z^3 (z-6 + \beta_z)
\eea
Note that in the important case where $z \rightarrow 2$ (and still $D=2$):
\bea
(z-2)\CL^{(4)}&=&\frac{(2-z)}{64}\left[3\left(\frac{\nabla_iN\nabla^iN}{N^2}\right)^2 -4 \frac{\nabla^i\nabla_iN}{N} \frac{\nabla_jN\nabla^jN}{N^2}\right].
\eea

A useful check is for $z=1$, which is the usual relativistic AdS case. The standard known result \cite{dhss} is that the 4th order term involving only spatial derivative is (up to total derivatives):
\bea
\CL^{(4)}&=&\left[\frac{1}{(D-3)(D-1)^2}\left(R_{\alpha\beta}R^{\alpha\beta}-\frac{D+1}{4D}R^2\right)\right]^{(4)}\cr
&=&\frac{1}{(D-3)(D-1)^2}\left[\left(\frac{\nabla^i\nabla_iN}{N}\right)^2+\left(\hat{R}_{ij}-\frac{\nabla_i\nabla_jN}{N}\right)\left(\hat{R}^{ij}-\frac{\nabla^i\nabla^jN}{N}\right)\right.\cr
&&\qquad\left.{}-\frac{D+1}{4D}\left(\hat{R}-\frac{2\nabla^i\nabla_iN}{N}\right)^2\right]\\
&=&-\frac{1}{(D-3)(D-1)^2}\left[-\left(\frac{\nabla_iN\nabla^iN}{N^2}\right)^2+ \frac{3\nabla_iN\nabla^iN\nabla^j\nabla_jN}{2N^3}+ \frac{\nabla^iN\nabla^jN \hat R_{ij}}{N^2}\right.\cr
&&\left.{}-\frac{1}{D}\frac{\nabla_i\nabla^iN}{N}\hat{R} -\frac{D-1}{D}\left(\frac{\nabla^i\nabla_iN}{N}\right)^2 -\hat{R}_{ij}\hat{R}^{ij}+\frac{D+1}{4D}\hat{R}^2\right]
\eea
This agrees exactly with the general result above. Of course, for $z=1$ there will also be contributing terms at this order which come from the $4z$ and $2+2z$ order terms (these will involve time derivatives).

An easily computable case is $D=1$ (for which $\hat{R}=0$). The above expressions yield $\displaystyle (z-3)\CL^{(4)}=\frac{(z-3)\nabla_iN\nabla^iN}{12z^3N^2}$. For $z=3$, which is when this would possibly generate a scaling anomaly, this expression vanishes.

\subsection{Four-derivative counterterms involving $\psi$}

There are many possible four-derivative counterterms involving $\psi$, for 
example:
\be
(z+D-4-\D_-)\CL^{(4+\D_-)}
=2K_{AB}^{(2)}\pi^{AB(2+\D_-)}+\pi_A^{(2)}\pi^{A(2+\D_-)}
+K_{AB}^{(\D_-)}T^{AB(4)}.
\ee
The right hand-side has been explicitly calculated and found to be zero in 
the case where $z=2$ and $D=2$.

\section{Anisotropic Weyl Anomaly in $2+1$ Dimensions}
\label{appwano}

Just as in the relativistic case, a theory which has the classical symmetry 
under anisotropic Weyl transformations can develop an anomaly in this symmetry 
at the quantum level.  Under the transformations
\be
\label{infaweyl}
\delta_\omega N = zN\delta\omega,\qquad\delta_\omega N_i = 2N_i\delta\omega,
\qquad\delta_\omega \gamma_{ij} = 2\gamma_{ij}\delta\omega,
\ee
the anomaly will show up as a nonvanishing variation of the partition 
function $\CZ[N,N_i,\gamma_{ij}]$, of the general form
\be
\delta_\omega \log\CZ[N,N_i,\gamma_{ij}] = -\int dt\,d^D\bx\, \sqrt{\gamma}\,N 
\,\FA\,\delta\omega,
\ee
where $\FA$ is now a function of $N$, $N_i$, and $\gamma_{ij}$.

We wish to determine what terms can arise in $\FA$.  As in the relativistic 
case, this question is cohomological in nature.%
\footnote{The cohomological approach to the relativistic Weyl anomaly was 
developed in \cite{bonora,bruno,deser}; see \cite{weinbk}, Chapter 22, 
for a general review of this approach.}
We introduce a 
nilpotent BRST operator $Q$, which acts on the metric multiplet via the 
infinitesimal anisotropic Weyl transformations (\ref{infaweyl}), with 
$\delta\omega$ replaced by a Grassmann parameter $c$ of ghost number one.    
We can represent this operator as
\be
Q= c\left(zN\frac{\delta}{\delta N}+2N_i \frac{\delta}{\delta N_i}+
2\gamma_{ij} \frac{\delta}{\delta \gamma_{ij}}\right).
\ee
Since $Q$ is nilpotent, the variation of the anomaly vanishes:
\be
Q \int dt\, d^D\bx\,\sqrt{\gamma}\,N\,\FA\, c  = -Q^2 \log\CZ = 0 .
\ee
This puts a constraint on the terms that can arise as $\FA$.

As usual, some of these terms can be removed by including appropriate 
counterterms.  If a term in the anomaly can be expressed as the variation 
some local counterterm, this (gravitational) counterterm can be subtracted 
from the action, thereby eliminating the associated anomaly.  Therefore the 
physical anomaly can be considered to lie in the cohomology of $Q$, 
at ghost number one.  The number of possible independent terms 
({\it i.e.}, generalized central charges) in the anomaly will be 
determined by the dimension of this cohomology.

In the case of $2+1$ dimensions with $z=2$, the anomaly must be -- on 
dimensional grounds -- a sum of terms of dimension four, lying in the 
cohomology of $Q$.  The list of possible terms is rather large; however, all 
but two are cohomologically trivial and can therefore be eliminated using 
local counterterms.  The only ones that cannot be removed are:
\be
\label{cohoq}
\hat K_{ij}\hat K^{ij}-\frac{1}{2}\hat K^2,
\qquad
\left\{\hat R -\left(\frac{\nabla N}{N}\right)^2 + \frac{\nabla^2 N}{N} 
\right\}^2.
\ee
We have seen in Section~\ref{bgmv} 
that the first cohomology class in (\ref{cohoq}), quadratic in 
the extrinsic curvature $\hat K_{ij}$, indeed arises in the holographic 
computation of the anisotropic 
Weyl anomaly, but the second one does not.  However, this term 
$\sim\hat R^2+\ldots$ is also non-trivial in the cohomology of $Q$, because 
$\sqrt{\gamma}\,N\,cR^2+\ldots$ cannot be obtained as the variation of another 
term (essentially  because variations of all available terms give rise to 
derivatives).  Hence, both classes should be expected to appear in the anomaly 
of generic $z=2$ field theories in $2+1$ dimensions.

In addition, we list $\FA$ for the five independent cocycles that contain 
only spatial derivatives, but are cohomologically trivial and can be 
eliminated by local counterterms: 
\be
\frac{1}{N}
\nabla^2\left[N\left(\hat R + \frac{\nabla^2 N}{N} 
- \left(\!\frac{\nabla N}{N}\!\right)^2\right)\right],
\ee
\be
\frac{1}{N}
\nabla_i\left(\left[\hat R + \frac{\nabla^2 N}{N} 
- \left(\!\frac{\nabla N}{N}\!\right)^2\right]\nabla^i N\right),
\ee
\be
\frac{1}{N}
\nabla^2\left[ \nabla^2 N - \frac{(\nabla N)^2}{N} \right],
\ee
\be
\frac{1}{N}
\nabla_i\left(\nabla^i N\left[\frac{\nabla^2 N}{N}-\left(\!
\frac{\nabla N}{N}\!\right)^2\right]-\frac{1}{2}\nabla^i\frac{(\nabla N)^2}{N}
\right),
\ee
\be
\frac{1}{N}
\nabla_i\left[\nabla^i N \left(\!\frac{\nabla N}{N}\!\right)^2\right].
\ee
This classification can be easily extended to include terms with time 
derivatives as well.  

As usual, this cohomology analysis only reveals the complete list of terms 
which {\it may\/} in principle occur in the anomaly.  Whether or not such 
terms are generated in a particular theory is a dynamical question, which 
requires an additional calculation.  

\bibliographystyle{JHEP}
\bibliography{lgh}

\providecommand{\href}[2]{#2}\begingroup\raggedright\begin{thebibliography}{10}

\bibitem{mqc}
P.~Ho\v{r}ava, {\it {Membranes at Quantum Criticality}},  {\em JHEP} {\bf 03}
  (2009) 020, [\href{http://xxx.lanl.gov/abs/arXiv:0812.4287}{{\tt
  arXiv:0812.4287}}].

\bibitem{lif}
P.~Ho\v{r}ava, {\it {Quantum Gravity at a Lifshitz Point}},  {\em Phys. Rev.}
  {\bf D79} (2009) 084008, [\href{http://xxx.lanl.gov/abs/arXiv:0901.3775}{{\tt
  arXiv:0901.3775}}].

\bibitem{grx}
P.~Ho\v{r}ava, {\it {General Covariance in Gravity at a Lifshitz Point}},  {\em
  Class. Quant. Grav.} {\bf 28} (2011) 114012,
  [\href{http://xxx.lanl.gov/abs/arXiv:1101.1081}{{\tt arXiv:1101.1081}}].

\bibitem{revmv}
M.~Visser, {\it {Status of Ho\v{r}ava gravity: A personal perspective}},  {\em
  J. Phys. Conf. Ser.} {\bf 314} (2011) 012002,
  [\href{http://xxx.lanl.gov/abs/arXiv:1103.5587}{{\tt arXiv:1103.5587}}].

\bibitem{revsm}
S.~Mukohyama, {\it {Ho\v{r}ava-Lifshitz Cosmology: A Review}},  {\em Class.
  Quant. Grav.} {\bf 27} (2010) 223101,
  [\href{http://xxx.lanl.gov/abs/arXiv:1007.5199}{{\tt arXiv:1007.5199}}].

\bibitem{ajld}
J.~Ambj{\o}rn, J.~Jurkiewicz, and R.~Loll, {\it {Spectral Dimension of the
  Universe}},  {\em Phys. Rev. Lett.} {\bf 95} (2005) 171301,
  [\href{http://xxx.lanl.gov/abs/hep-th/0505113}{{\tt hep-th/0505113}}].

\bibitem{ajl}
J.~Ambj{\o}rn, J.~Jurkiewicz, and R.~Loll, {\it {Quantum Gravity as Sum over
  Spacetimes}},  {\em Lect. Notes Phys.} {\bf 807} (2010) 59--124,
  [\href{http://xxx.lanl.gov/abs/0906.3947}{{\tt 0906.3947}}].

\bibitem{dario}
D.~Benedetti and J.~Henson, {\it {Spectral Geometry as a Probe of Quantum
  Spacetime}},  {\em Phys.Rev.} {\bf D80} (2009) 124036,
  [\href{http://xxx.lanl.gov/abs/0911.0401}{{\tt 0911.0401}}].

\bibitem{wdavis}
C.~Anderson, S.~Carlip, J.~H. Cooperman, P.~Ho\v{r}ava, R.~Kommu, and P.~R.
  Zulkowski, {\it {Quantizing Ho\v{r}ava-Lifshitz Gravity via Causal Dynamical
  Triangulations}},  \href{http://xxx.lanl.gov/abs/arXiv:1111.6634}{{\tt
  arXiv:1111.6634}}.

\bibitem{revsh}
S.~A. Hartnoll, {\it {Lectures on Holographic Methods for Condensed Matter
  Physics}},  {\em Class. Quant. Grav.} {\bf 26} (2009) 224002,
  [\href{http://xxx.lanl.gov/abs/arXiv:0903.3246}{{\tt arXiv:0903.3246}}].

\bibitem{revmcg}
J.~McGreevy, {\it {Holographic Duality with a View toward Many-Body Physics}},
  {\em Adv. High Energy Phys.} {\bf 2010} (2010) 723105,
  [\href{http://xxx.lanl.gov/abs/arXiv:0909.0518}{{\tt arXiv:0909.0518}}].

\bibitem{revss}
S.~Sachdev, {\it {What Can Gauge-Gravity Duality Teach us about Condensed
  Matter Physics?}},  \href{http://xxx.lanl.gov/abs/arXiv:1108.1197}{{\tt
  arXiv:1108.1197}}.

\bibitem{vhmr}
V.~E. Hubeny and M.~Rangamani, {\it {A Holographic View on Physics out of
  Equilibrium}},  {\em Adv.High Energy Phys.} {\bf 2010} (2010) 297916,
  [\href{http://xxx.lanl.gov/abs/arXiv:1006.3675}{{\tt arXiv:1006.3675}}].

\bibitem{son}
D.~T. Son, {\it {Toward an AdS/Cold Atoms Correspondence: A Geometric
  Realization of the Schr\"{o}dinger Symmetry}},  {\em Phys. Rev.} {\bf D78}
  (2008) 046003, [\href{http://xxx.lanl.gov/abs/arXiv:0804.3972}{{\tt
  arXiv:0804.3972}}].

\bibitem{bmcg}
K.~Balasubramanian and J.~McGreevy, {\it {Gravity Duals for Non-Relativistic
  CFTs}},  {\em Phys. Rev. Lett.} {\bf 101} (2008) 061601,
  [\href{http://xxx.lanl.gov/abs/arXiv:0804.4053}{{\tt arXiv:0804.4053}}].

\bibitem{klm}
S.~Kachru, X.~Liu, and M.~Mulligan, {\it {Gravity Duals of Lifshitz-like Fixed
  Points}},  {\em Phys.Rev.} {\bf D78} (2008) 106005,
  [\href{http://xxx.lanl.gov/abs/arXiv:0808.1725}{{\tt arXiv:0808.1725}}].

\bibitem{penrin}
R.~Penrose and W.~Rindler, {\em {Spinors and Space-Time, Vol. 2}}.
\newblock Cambridge Univ. Press, 1986.

\bibitem{aci}
P.~Ho\v{r}ava and C.~M. Melby-Thompson, {\it {Anisotropic Conformal Infinity}},
   {\em Gen. Rel. Grav.} {\bf 43} (2010) 1391,
  [\href{http://xxx.lanl.gov/abs/arXiv:0909.3841}{{\tt arXiv:0909.3841}}].

\bibitem{mans}
M.~Henningson and K.~Skenderis, {\it {The Holographic Weyl anomaly}},  {\em
  JHEP} {\bf 9807} (1998) 023,
  [\href{http://xxx.lanl.gov/abs/arXiv:hep-th/9806087}{{\tt
  arXiv:hep-th/9806087}}].

\bibitem{weylduff}
M.~Duff, {\it {Twenty years of the Weyl anomaly}},  {\em Class. Quant. Grav.}
  {\bf 11} (1994) 1387--1404,
  [\href{http://xxx.lanl.gov/abs/arXiv:hep-th/9308075}{{\tt
  arXiv:hep-th/9308075}}].

\bibitem{bgmr}
V.~Balasubramanian, E.~G. Gimon, D.~Minic, and J.~Rahmfeld, {\it
  {Four-dimensional conformal supergravity from AdS space}},  {\em Phys. Rev.}
  {\bf D63} (2001) 104009,
  [\href{http://xxx.lanl.gov/abs/arXiv:hep-th/0007211}{{\tt
  arXiv:hep-th/0007211}}].

\bibitem{confsg}
E.~Fradkin and A.~A. Tseytlin, {\it {Conformal Supergravity}},  {\em Phys.
  Rept.} {\bf 119} (1985) 233--362.

\bibitem{theisen}
I.~Adam, I.~V. Melnikov, and S.~Theisen, {\it {A Non-Relativistic Weyl
  Anomaly}},  {\em JHEP} {\bf 0909} (2009) 130,
  [\href{http://xxx.lanl.gov/abs/arXiv:0907.2156}{{\tt arXiv:0907.2156}}].

\bibitem{ibdl}
I.~Bakas and D.~{L\"{u}st}, {\it {Axial Anomalies of Lifshitz Fermions}},
  \href{http://xxx.lanl.gov/abs/arXiv:1103.5693}{{\tt arXiv:1103.5693}}.

\bibitem{ib}
I.~Bakas, {\it {More on Axial Anomalies of Lifshitz Fermions}},
  \href{http://xxx.lanl.gov/abs/arXiv:1110.1332}{{\tt arXiv:1110.1332}}.

\bibitem{revsk}
K.~Skenderis, {\it {Lecture notes on holographic renormalization}},  {\em
  Class. Quant. Grav.} {\bf 19} (2002) 5849,
  [\href{http://xxx.lanl.gov/abs/arXiv:hep-th/0209067}{{\tt
  arXiv:hep-th/0209067}}].

\bibitem{malda}
J.~Maldacena, {\it {Einstein Gravity from Conformal Gravity}},
  \href{http://xxx.lanl.gov/abs/arXiv:1105.5632}{{\tt arXiv:1105.5632}}.

\bibitem{maldai}
J.~M. Maldacena, {\it {Non-Gaussian features of primordial fluctuations in
  single field inflationary models}},  {\em JHEP} {\bf 0305} (2003) 013,
  [\href{http://xxx.lanl.gov/abs/astro-ph/0210603}{{\tt astro-ph/0210603}}].

\bibitem{harlow}
D.~Harlow and D.~Stanford, {\it {Operator Dictionaries and Wave Functions in
  AdS/CFT and dS/CFT}},  \href{http://xxx.lanl.gov/abs/arXiv:1104.2621}{{\tt
  arXiv:1104.2621}}.

\bibitem{mcf1}
P.~McFadden and K.~Skenderis, {\it {Holography for Cosmology}},  {\em
  Phys.Rev.} {\bf D81} (2010) 021301,
  [\href{http://xxx.lanl.gov/abs/arXiv:0907.5542}{{\tt arXiv:0907.5542}}].

\bibitem{mcf2}
P.~McFadden and K.~Skenderis, {\it {Cosmological 3-point Correlators from
  Holography}},  {\em JCAP} {\bf 1106} (2011) 030,
  [\href{http://xxx.lanl.gov/abs/arXiv:1104.3894}{{\tt arXiv:1104.3894}}].

\bibitem{mcf3}
A.~Bzowski, P.~McFadden, and K.~Skenderis, {\it {Holographic Predictions for
  Cosmological 3-point Functions}},  {\em JHEP} {\bf 1203} (2012) 091,
  [\href{http://xxx.lanl.gov/abs/arXiv:1112.1967}{{\tt arXiv:1112.1967}}].

\bibitem{gen}
P.~Ho\v{r}ava and C.~M. Melby-Thompson, {\it {General Covariance in Quantum
  Gravity at a Lifshitz Point}},  {\em Phys. Rev.} {\bf D82} (2010) 064027,
  [\href{http://xxx.lanl.gov/abs/arXiv:1007.2410}{{\tt arXiv:1007.2410}}].

\bibitem{bpstwo}
D.~Blas, O.~Pujol\`{a}s, and S.~Sibiryakov, {\it {Consistent Extension of
  Ho\v{r}ava Gravity}},  {\em Phys. Rev. Lett.} {\bf 104} (2010) 181302,
  [\href{http://xxx.lanl.gov/abs/arXiv:0909.3525}{{\tt arXiv:0909.3525}}].

\bibitem{bpsthree}
D.~Blas, O.~Pujol\`{a}s, and S.~Sibiryakov, {\it {Models of non-relativistic
  quantum gravity: The Good, the bad and the healthy}},  {\em JHEP} {\bf 1104}
  (2011) 018, [\href{http://xxx.lanl.gov/abs/arXiv:1007.3503}{{\tt
  arXiv:1007.3503}}].

\bibitem{taylor}
M.~Taylor, {\it {Non-relativistic Holography}},
  \href{http://xxx.lanl.gov/abs/arXiv:0812.0530}{{\tt arXiv:0812.0530}}.

\bibitem{st1}
K.~Balasubramanian and K.~Narayan, {\it {Lifshitz Spacetimes from AdS Null and
  Cosmological Solutions}},  {\em JHEP} {\bf 1008} (2010) 014,
  [\href{http://xxx.lanl.gov/abs/arXiv:1005.3291}{{\tt arXiv:1005.3291}}].

\bibitem{st2}
A.~Donos and J.~P. Gauntlett, {\it {Lifshitz Solutions of D=10 and D=11
  Supergravity}},  {\em JHEP} {\bf 1012} (2010) 002,
  [\href{http://xxx.lanl.gov/abs/arXiv:1008.2062}{{\tt arXiv:1008.2062}}].

\bibitem{st3}
R.~Gregory, S.~L. Parameswaran, G.~Tasinato, and I.~Zavala, {\it {Lifshitz
  Solutions in Supergravity and String Theory}},  {\em JHEP} {\bf 1012} (2010)
  047, [\href{http://xxx.lanl.gov/abs/arXiv:1009.3445}{{\tt arXiv:1009.3445}}].

\bibitem{st4}
A.~Donos, J.~P. Gauntlett, N.~Kim, and O.~Varela, {\it {Wrapped M5-Branes,
  Consistent Truncations and AdS/CMT}},  {\em JHEP} {\bf 1012} (2010) 003,
  [\href{http://xxx.lanl.gov/abs/arXiv:1009.3805}{{\tt arXiv:1009.3805}}].

\bibitem{hawkinge}
S.~W. Hawking and G.~F.~R. Ellis, {\em {The Large Scale Structure of
  Space-time}}.
\newblock Cambridge Univ. Press, 1973.

\bibitem{ross}
S.~F. Ross, {\it {Holography for asymptotically locally Lifshitz spacetimes}},
  {\em Class. Quant. Grav.} {\bf 28} (2011) 215019,
  [\href{http://xxx.lanl.gov/abs/arXiv:1107.4451}{{\tt arXiv:1107.4451}}].

\bibitem{tasim}
J.~M. Maldacena, {\it {TASI 2003 Lectures on AdS/CFT}},
  \href{http://xxx.lanl.gov/abs/arXiv:hep-th/0309246}{{\tt
  arXiv:hep-th/0309246}}.

\bibitem{tasip}
J.~Polchinski, {\it {Introduction to Gauge/Gravity Duality}},
  \href{http://xxx.lanl.gov/abs/arXiv:1010.6134}{{\tt arXiv:1010.6134}}.

\bibitem{vijayper}
V.~Balasubramanian and P.~Kraus, {\it {A Stress Tensor for Anti-de Sitter
  Gravity}},  {\em Commun. Math. Phys.} {\bf 208} (1999) 413--428,
  [\href{http://xxx.lanl.gov/abs/arXiv:hep-th/9902121}{{\tt
  arXiv:hep-th/9902121}}].

\bibitem{perfinn}
P.~Kraus, F.~Larsen, and R.~Siebelink, {\it {The Gravitational Action in
  Asymptotically AdS and Flat Spacetimes}},  {\em Nucl. Phys.} {\bf B563}
  (1999) 259--278, [\href{http://xxx.lanl.gov/abs/arXiv:hep-th/9906127}{{\tt
  arXiv:hep-th/9906127}}].

\bibitem{dhss}
S.~de~Haro, S.~N. Solodukhin, and K.~Skenderis, {\it {Holographic
  Reconstruction of Spacetime and Renormalization in the AdS/CFT
  Correspondence}},  {\em Commun. Math. Phys.} {\bf 217} (2001) 595--622,
  [\href{http://xxx.lanl.gov/abs/arXiv:hep-th/0002230}{{\tt
  arXiv:hep-th/0002230}}].

\bibitem{bfs}
M.~Bianchi, D.~Z. Freedman, and K.~Skenderis, {\it {Holographic
  Renormalization}},  {\em Nucl.Phys.} {\bf B631} (2002) 159--194,
  [\href{http://xxx.lanl.gov/abs/arXiv:hep-th/0112119}{{\tt
  arXiv:hep-th/0112119}}].

\bibitem{dbvv}
J.~de~Boer, E.~P. Verlinde, and H.~L. Verlinde, {\it {On the Holographic
  Renormalization Group}},  {\em JHEP} {\bf 0008} (2000) 003,
  [\href{http://xxx.lanl.gov/abs/arXiv:hep-th/9912012}{{\tt
  arXiv:hep-th/9912012}}].

\bibitem{revdb}
J.~de~Boer, {\it {The Holographic renormalization group}},  {\em Fortsch.Phys.}
  {\bf 49} (2001) 339--358,
  [\href{http://xxx.lanl.gov/abs/arXiv:hep-th/0101026}{{\tt
  arXiv:hep-th/0101026}}].

\bibitem{revjp}
M.~Fukuma, S.~Matsuura, and T.~Sakai, {\it {Holographic renormalization
  group}},  {\em Prog. Theor. Phys.} {\bf 109} (2003) 489,
  [\href{http://xxx.lanl.gov/abs/arXiv:hep-th/0212314}{{\tt
  arXiv:hep-th/0212314}}].

\bibitem{mann}
R.~Mann and R.~McNees, {\it {Holographic Renormalization for Asymptotically
  Lifshitz Spacetimes}},  {\em JHEP} {\bf 1110} (2011) 129,
  [\href{http://xxx.lanl.gov/abs/arXiv:1107.5792}{{\tt arXiv:1107.5792}}].

\bibitem{deboer}
M.~Baggio, J.~de~Boer, and K.~Holsheimer, {\it {Hamilton-Jacobi Renormalization
  for Lifshitz Spacetime}},
  \href{http://xxx.lanl.gov/abs/arXiv:1107.5562}{{\tt arXiv:1107.5562}}.

\bibitem{papask}
I.~Papadimitriou and K.~Skenderis, {\it {AdS / CFT correspondence and
  geometry}},  \href{http://xxx.lanl.gov/abs/arXiv:hep-th/0404176}{{\tt
  arXiv:hep-th/0404176}}.

\bibitem{papaske}
I.~Papadimitriou and K.~Skenderis, {\it {Correlation functions in holographic
  RG flows}},  {\em JHEP} {\bf 0410} (2004) 075,
  [\href{http://xxx.lanl.gov/abs/arXiv:hep-th/0407071}{{\tt
  arXiv:hep-th/0407071}}].

\bibitem{ffg}
C.~Fefferman and C.~R. Graham, {\it {Conformal Invariants, in: Elie Cartan et
  les Math\'{e}matiques d'aujourd'hui}},  {\em Ast\'{e}risque} (1985) 95.

\bibitem{graham}
C.~R. Graham, {\it {Volume and Area Renormalizations for Conformally Compact
  Einstein Metrics}},  \href{http://xxx.lanl.gov/abs/arXiv:math/9909042}{{\tt
  arXiv:math/9909042}}.

\bibitem{gw}
C.~R. Graham and E.~Witten, {\it {Conformal Anomaly of Submanifold Observables
  in AdS/CFT Correspondence}},  {\em Nucl. Phys.} {\bf B546} (1999) 52--64,
  [\href{http://xxx.lanl.gov/abs/arXiv:hep-th/9901021}{{\tt
  arXiv:hep-th/9901021}}].

\bibitem{sotdb}
D.~Vernieri and T.~P. Sotiriou, {\it {Ho\v{r}ava-Lifshitz Gravity: Detailed
  Balance Revisited}},  \href{http://xxx.lanl.gov/abs/arXiv:1112.3385}{{\tt
  arXiv:1112.3385}}.

\bibitem{phroma}
P.~Ho\v{r}ava, {\it {Gravity at a Lifshitz Point}},  {\em {\rm review talk at}
  Strings 2009} (Roma, Italy) June 25, 2009 (unpublished),
  [\href{http://xxx.lanl.gov/abs/http://strings2009.roma2.infn.it/talks/Horava%
\_Strings09.pdf}{{\tt
  http://strings2009.roma2.infn.it/talks/Horava\_Strings09.pdf}}].

\bibitem{ghmt}
K.~T. Grosvenor, P.~Ho\v{r}ava, and C.~M. Melby-Thompson, {\it {Quantum Gravity
  with Anisotropic Scaling Near the Schwarzschild Horizon}},  {\em {\rm to
  appear}}.

\bibitem{eme}
C.~Xu and P.~Ho\v{r}ava, {\it {Emergent Gravity at a Lifshitz Point from a Bose
  Liquid on the Lattice}},  {\em Phys. Rev.} {\bf D81} (2010) 104033,
  [\href{http://xxx.lanl.gov/abs/arXiv:1003.0009}{{\tt arXiv:1003.0009}}].

\bibitem{grisha}
G.~Volovik, {\it {Topology of Quantum Vacuum}},
  \href{http://xxx.lanl.gov/abs/arXiv:1111.4627}{{\tt arXiv:1111.4627}}.

\bibitem{omid}
S.~F. Ross and O.~Saremi, {\it {Holographic stress tensor for non-relativistic
  theories}},  {\em JHEP} {\bf 0909} (2009) 009,
  [\href{http://xxx.lanl.gov/abs/arXiv:0907.1846}{{\tt arXiv:0907.1846}}].

\bibitem{bruths}
H.~Braviner, R.~Gregory, and S.~F. Ross, {\it {Flows Involving Lifshitz
  Solutions}},  {\em Class. Quant. Grav.} {\bf 28} (2011) 225028,
  [\href{http://xxx.lanl.gov/abs/arXiv:1108.3067}{{\tt arXiv:1108.3067}}].

\bibitem{xi}
M.~C. Cheng, S.~A. Hartnoll, and C.~A. Keeler, {\it {Deformations of Lifshitz
  holography}},  {\em JHEP} {\bf 1003} (2010) 062,
  [\href{http://xxx.lanl.gov/abs/arXiv:0912.2784}{{\tt arXiv:0912.2784}}].

\bibitem{bonora}
L.~Bonora, P.~Pasti, and M.~Bregola, {\it {Weyl Cocycles}},  {\em Class. Quant.
  Grav.} {\bf 3} (1986) 635.

\bibitem{bruno}
W.~A. Bardeen and B.~Zumino, {\it {Consistent and Covariant Anomalies in Gauge
  and Gravitational Theories}},  {\em Nucl. Phys.} {\bf B244} (1984) 421.

\bibitem{deser}
S.~Deser and A.~Schwimmer, {\it {Geometric classification of conformal
  anomalies in arbitrary dimensions}},  {\em Phys. Lett.} {\bf B309} (1993)
  279, [\href{http://xxx.lanl.gov/abs/arXiv:hep-th/9302047}{{\tt
  arXiv:hep-th/9302047}}].

\bibitem{weinbk}
S.~Weinberg, {\em {The Quantum Theory of Fields, Vol. 2}}.
\newblock Cambridge Univ. Press, 1996.

\end{thebibliography}\endgroup
\end{document}